\documentclass[12pt]{article}
\usepackage[paper=a4paper,margin=1in]{geometry}
\usepackage{t1enc}      

\usepackage{amsmath}
\usepackage{amssymb}

\newcommand{\edth} {\mbox{\symbol{'360}}}

\newcommand{\uA}{\underline A \,}
\newcommand{\uB}{\underline B \,}

\newcommand{\bi}{\bf i}
\newcommand{\bj}{\bf j}
\newcommand{\bk}{\bf k}

\numberwithin{equation}{section}

\begin{document}
\bibliographystyle{unsrt}

\title{Witten spinors on maximal, conformally flat hypersurfaces}

\author{J\"org Frauendiener\\
  Department of Mathematics and Statistics, University of Otago, \\
  P. O. Box 56, Dunedin 9010, New Zealand
\and
James M. Nester \\
  Department of Physics, Graduate Institute of Astronomy, \\
  and Center for Mathematics and Theoretical Physics,\\
  National Central University, Chungli 320, Taiwan
\and
L\'aszl\'o B. Szabados \\
  Research Institute for Particle and Nuclear Physics, \\
  H--1525 Budapest 114, P. O. Box 49, Hungary}

\maketitle

\begin{abstract}
The boundary conditions that exclude zeros of the solutions of the 
Witten equation (and hence guarantee the existence of a 3-frame 
satisfying the so-called special orthonormal frame gauge conditions) 
are investigated. We determine the general form of the conformally 
invariant boundary conditions for the Witten equation, and find the 
boundary conditions that characterize the constant and the 
conformally constant spinor fields among the solutions of the 
Witten equations on compact domains in extrinsically and 
intrinsically flat, and on maximal, intrinsically globally 
conformally flat spacelike hypersurfaces, respectively. 

We also provide a number of exact solutions of the Witten equation 
with various boundary conditions (both at infinity and on inner or 
outer boundaries) that single out nowhere vanishing spinor fields 
on the flat, non-extreme Reissner--Nordstr\"om and Brill--Lindquist 
data sets. Our examples show that there is an interplay between the 
boundary conditions, the global topology of the hypersurface and 
the existence/non-existence of zeros of the solutions of the Witten 
equation. 

\end{abstract}


\section{Introduction}
\label{sec-1}

It is well known that spinorial techniques, and in particular the 
Witten equation and the use of two-component spinors, greatly 
simplified the proof of the positivity of both the ADM and 
Bondi--Sachs energies, even in the presence of black holes (see e.g. 
\cite{GH,RT}). However, spinors and the Witten equation play only an 
auxiliary role in the proofs. The only essential point in these 
investigations was the {\em existence} of the solution with given 
asymptotic behaviour. The spinor fields could have zeros, and the 
zeros did not have any significance. 

A slightly different approach to the proof of the positivity of the 
gravitational energy was suggested in \cite{Ne}. That was based on 
the tetrad formulation of general relativity, and to obtain a simple 
positivity proof a {\em gauge condition} for the orthonormal frame 
fields had to be imposed \cite{N1,N2}. However, since there is a 
natural correspondence between {\em nowhere vanishing} spinor fields 
on a spacelike hypersurface and {\em non-singular} orthogonal vector 
bases (triads) there \cite{J}, one could expect that the orthonormal 
frame gauge condition could be translated into the language of 
spinors. Indeed, this gauge condition has been reformulated in this 
way, and finding its solution is equivalent to finding a solution 
of the Witten equation with vanishing extrinsic curvature but, in 
general, with a mass term \cite{DMH}. Thus the great advantage of 
the use of spinors is that while the frame gauge condition in its 
original form is a system of {\em non-linear} elliptic partial 
differential equations, in its spinorial form it is the {\em linear} 
Witten equation. 

However, the solution of the Witten equation can have zeros, which 
can even form a two-codimensional set \cite{bar}. Thus, to have a 
perfectly well defined frame field on a given domain, the spinor 
field should have no zero there. In \cite{Ne07}, it was argued that 
while spinor zeros could occur they are not {\em generic\/}, but no 
general guidance was given there as to how to know when they would 
or would not occur. Since the Witten (or, more generally, a Dirac 
type) equation is elliptic, its solution depends also on the boundary 
condition in an essential way. Hence the frame gauge condition 
consists not only of the elliptic differential equation on the 
spacelike hypersurface, but an appropriate boundary condition 
selecting a non-degenerate solution must also be specified. Thus, in 
the spinorial language, the question is: How to find the boundary 
conditions for the Witten equation that ensure the existence of a 
nowhere vanishing solution on the given domain? 

The existence of the global solution of the frame gauge condition 
could be a very useful tool in various problems in general relativity. 
In fact, this could provide a geometrically preferred rigid system of 
frames of reference on an extended domain of a spacelike hypersurface, 
and the frame would be controlled only by appropriate boundary 
conditions. Thus from the point of view of applications, it is 
desirable to find the conditions that guarantee the existence of 
globally non-singular solutions of the frame gauge conditions, or, in 
other words, the existence of nowhere vanishing solutions of the 
Witten equation.

Unfortunately, the general problem of finding the appropriate boundary 
conditions appears to be surprisingly difficult. On the other hand, 
there are still physically important special cases in which there is 
some hope of being able to clarify the boundary conditions for nowhere 
vanishing solutions. Such are the maximal, globally conformally flat 
data sets, which include e.g. the Reissner--Nordstr\"om \cite{Gi}, 
Brill--Lindquist \cite{BL} and Bowen--York \cite{BoY} data sets. These 
data sets represent finitely many black holes with specified total 
mass and linear and angular momenta at spatial infinity. 

If the spacelike hypersurface is maximal and intrinsically {\em flat}, 
e.g. when the Cauchy data induced on the spacelike hypersurface are 
flat (such as a spacelike hyperplane in Minkowski spacetime), then 
the gauge condition is expected to yield a {\em constant} spinor 
field. Thus the boundary condition in the general, curved case must 
have a form that reduces to the one specifying the constant spinor 
field in the flat case. In fact, in the flat case it is natural to 
expect the geometrically preferred orthonormal frame fields to be 
just the Cartesian ones, which are in a one-to-one correspondence 
with the constant spinor fields up to a real constant scale factor. 
This raises the question of whether it is natural to expect that on 
intrinsically {\em conformally flat} hypersurfaces the spinor field 
should be proportional to a spinor field which is constant with 
respect to the flat connection, and if that is the case, then what is 
the appropriate boundary condition?

Here we investigate the question of the boundary conditions for the 
Witten equation that yield a nowhere vanishing spinor field in the 
special case when $\Sigma$ is maximal and its intrinsic geometry is 
globally conformally flat. We determine the boundary conditions 
that yield the {\em conformally constant}, and hence nowhere 
vanishing spinor fields on compact domains in $\Sigma$. Thus, in 
particular, the frame gauge condition can be satisfied on such 
domains in maximal, globally conformally flat spacelike 
hypersurfaces. On the other hand, as the examples of the 
non-extreme Reissner-Nordstr\"om data set show, the boundary 
conditions on the different connected components of the boundary {\em 
at infinity} cannot be chosen independently. 

In the next section, we recall the Witten equation (both in its 
covariant form and in the Geroch--Held--Penrose (GHP) formalism) 
and clarify its conformal properties and the relationship between 
the solutions of the Witten equations and certain geometrically 
distinguished orthogonal vector bases. Then we determine the general 
form of the conformally invariant boundary conditions for the Witten 
equation. 

In section \ref{sec-3}, explicit solutions of the Witten equation 
are given. First, in subsection \ref{sub-3.1}, a simple solution is 
given which illustrates what kinds of zeros may appear. Then, in 
subsection \ref{sub-3.2}, we determine the fundamental (in some sense 
spherically symmetric) solutions of the Witten equation on 
intrinsically and extrinsically flat 3-spaces. We also discuss the 
potentially reasonable explicitly given boundary conditions both at 
infinity and at boundaries that are metric spheres of finite radius. 
We will see that several apparently natural boundary conditions that 
have already appeared in various problems can yield spinor fields with 
one or more zeros. 
Subsection \ref{sub-3.3} is devoted to the solution of the Witten 
equation on maximal, globally conformally flat spacelike hypersurfaces, 
namely on the complete, non-extremal Reissner--Nordstr\"om and 
Brill--Lindquist data sets, as well as on a part of the 
Reissner--Nordstr\"om data set with inner boundary. We will see that 
on the complete Reissner--Nordstr\"om data set, the asymptotic values 
of the nowhere vanishing solutions of the Witten equation cannot be 
chosen independently on the two infinities. 

Finally, in section \ref{sec-4} we determine the boundary conditions 
on general boundaries in terms of the boundary Dirac-operator (built 
from a Sen type connection) that single out the {\em constant} and 
the {\em conformally constant} spinor fields from all the possible 
solutions of the Witten equation on compact domains in flat and in 
maximal, globally conformally flat spacelike hypersurfaces on which 
the Hamiltonian constraint equation is satisfied and the weak energy 
condition holds, respectively. These imply, in particular, that in 
these situations the orthonormal frame gauge condition can be used 
as a perfect gauge condition. Our results are summarized in section 
\ref{sec-5}. 

Our conventions and notations are mostly those of \cite{PR,HT}. 
In particular, we use the abstract index formalism, and only the 
boldface and underlined capital Latin indices are concrete name 
indices. Abstract Lorentzian tensor indices are freely converted 
to pairs of unprimed and primed spinor indices and back, according 
to the rule $a\mapsto AA'$. The signature of the spacetime metric 
is $-2$, and the spacetime Riemann and Ricci tensors, and the 
scalar curvature are defined by $-{}^4R^a{}_{bcd}X^bY^cZ^d:=\nabla
_Y(\nabla_ZX^a)-\nabla_Z(\nabla_YX^a)-\nabla_{[Y,Z]}X^a$, ${}^4R_{ab}
:={}^4R^c{}_{acb}$ and ${}^4R:={}^4R_{ab}g^{ab}$, respectively, and 
hence Einstein's equations take the form ${}^4G_{ab}=-\kappa T
_{ab}$, $\kappa>0$. We use the standard definitions for the GHP 
spin coefficients and the edth operators given in \cite{PR}.


\section{Boundary conditions for the Witten equation}
\label{sec-2}

\subsection{The Witten equation}
\label{sub-2.1}

\subsubsection{The covariant form}
\label{sub-2.1.1}

Let $\Sigma$ be a smooth spacelike hypersurface of a space-time $(M,
g_{ab})$, $h_{ab}$ the induced (negative definite) metric and $\chi
_{ab}$ the extrinsic curvature on $\Sigma$. Let $t^a$ be the future 
pointing unit timelike normal of $\Sigma$, and define $P^a_b:=\delta
^a_b-t^at_b$, the orthogonal projection to $\Sigma$. If $D_a$ denotes 
the intrinsic Levi-Civita derivative operator on $\Sigma$ determined 
by $h_{ab}$ and ${\cal D}_a:=P^b_a\nabla_b$, the so-called Sen 
operator \cite{Se}, then 
\begin{equation}
{\cal D}_{A'A}\lambda^A=D_{A'A}\lambda^A+\frac{1}{2}\chi t_{A'A}
\lambda^A, \label{eq:2.1.1}
\end{equation}
and the Witten equation is ${\cal D}_{A'A}\lambda^A=0$. Here $\chi$
is the trace of the extrinsic curvature. 

Next we clarify the behaviour of (\ref{eq:2.1.1}) under the conformal 
rescaling of the {\em spacetime metric} $g_{ab}$ by the conformal 
factor $\Omega$. Suppose that $\lambda_A$ has the conformal weight 
$w$, i.e. under such a conformal rescaling it transforms as $\lambda
_A\mapsto\hat\lambda_A:=\Omega^w\lambda_A$. Then, since under such a 
conformal rescaling the spinor derivative is well known \cite{PR,HT} 
to transform as $\nabla_{AA'}\lambda^B\mapsto\hat\nabla_{AA'}\lambda
^B=\nabla_{AA'}\lambda^B+\delta^B_A\lambda^C\Upsilon_{CA'}$, where 
$\Upsilon_e:=\nabla_e\ln\Omega$, it is easy to deduce that 
\begin{equation}
\Omega^{w-1}{\cal D}_{A'A}\lambda^A=\hat D_{A'A}\hat\lambda^A+
\frac{1}{2}\chi t_{A'A}\hat\lambda^A-(w+\frac{1}{2})\hat\lambda^A
\hat D_{AA'}\ln\Omega.  \label{eq:2.1.2}
\end{equation}
(N.B.: Under the conformal rescaling above, the extrinsic curvature 
transforms as $\chi_{ab}\mapsto\Omega\chi_{ab}+t^e(\nabla_e\Omega)h
_{ab}$, and hence for the transformation of the mean curvature we 
obtain $\chi\mapsto\hat\chi=\Omega^{-1}\chi+3\Omega^{-1}t^e\Upsilon
_e$.) Therefore, on {\em maximal, intrinsically globally conformally 
flat} hypersurfaces the solution $\lambda^A$ of the Witten equation 
can be recovered as $\lambda^A=\Omega^{\frac{3}{2}}\hat\lambda^A$ 
from the solution of the {\em intrinsically and extrinsically flat} 
Witten equation $\hat D_{A'A}\hat\lambda^A=0$. 
Thus, it is natural to assign the conformal weight $w=-\frac{1}{2}$ 
to $\lambda_A$. If, for the sake of simplicity, the conformal 
rescaling is chosen to be {\em purely spatial}, i.e. $t^e\nabla_e
\Omega=0$ (which will be done in what follows), then the rescaling 
preserves the maximality of the hypersurface.

\subsubsection{The GHP form}
\label{sub-2.1.2}

Next suppose that the spacelike hypersurface $\Sigma$ is foliated 
by 2-surfaces ${\cal S}_r$ of spherical topology. Let $v^e$ denote 
their outward pointing unit spacelike normal tangent to $\Sigma$, 
and introduce the projection $\Pi^a_b:=P^a_b+v^av_b$ to the surfaces 
${\cal S}_r$. Then we can decompose ${\cal D}_{A'A}\lambda^A$ with 
respect to this foliation as 
\begin{equation}
{\cal D}_{A'A}\lambda^A=\bigl(\Pi^{B'B}_{A'A}-v^{B'B}v_{A'A}\bigr)
{\cal D}_{B'B}\lambda^A=\Delta_{A'A}\lambda^A-v^e\bigl({\cal D}_e
\lambda^A\bigr)v_{AA'}, \label{eq:2.1.3}
\end{equation}
where $\Delta_e:=\Pi^f_e\nabla_f$ is a Sen type derivative operator 
on the surfaces ${\cal S}_r$. (Note that this derivative operator 
deviates from the intrinsic Levi-Civita derivative operator $\delta
_e$ by the extrinsic curvature tensor of ${\cal S}_r$ \cite{Sz1}.) 
The contraction $\Delta_{A'A}\lambda^A$ is only a part of the 
derivative $\Delta_{A'A}\lambda_B$. The remaining part is 
represented by 
\begin{equation*}
{\cal T}_{A'AB}{}^C\lambda_C:=\Delta_{A'(A}\lambda_{B)}+\frac{1}{2}
\gamma_{AB}\gamma^{CD}\Delta_{A'C}\lambda_D,
\end{equation*}
where $\gamma^A{}_B:=2t^{AA'}v_{BA'}$. (N.B.: $\gamma^A{}_B$ is 
independent of the actual choice for the two normals $t^a$ and $v^a$, 
it is completely determined by ${\cal S}_r$.) This ${\cal T}_{A'AB}
{}^C\lambda_C$ is just the covariant form of the 2-surface twistor 
operator on the 2-surfaces ${\cal S}_r$, and the 2-surface twistor 
equation is ${\cal T}_{A'AB}{}^C\lambda_C=0$. This is, in fact, two 
equations (see below and \cite{PR,HT,Sz1}). 

Let us introduce a GHP spin frame field $(o^A,\iota^A)$ on $\Sigma$ 
such that 
\begin{equation}
t^{AA'}=\frac{1}{2}o^A\bar o^{A'}+\iota^A\bar\iota^{A'},
\hskip 20pt
v^{AA'}=\frac{1}{2}o^A\bar o^{A'}-\iota^A\bar\iota^{A'};
\label{eq:2.1.4}
\end{equation}
and define the spinor components in this basis by $\lambda_0:=
\lambda_Ao^A$ and $\lambda_1:=\lambda_A\iota^A$. Then in the GHP 
formalism, the Witten equation ${\cal D}_{A'A}\lambda^A=0$ takes 
the form 
\begin{eqnarray}
0\!\!\!\!&=\!\!\!\!&{\edth}'\lambda_0-v^e\partial_e\lambda_1+\Bigl(
 \rho-v^e\bigl({\cal D}_e\iota_A\bigr)o^A\Bigr)\lambda_1+v^e\bigl(
 {\cal D}_e\iota_A\bigr)\iota^A\lambda_0, \label{eq:2.1.5a} \\
0\!\!\!\!&=\!\!\!\!&{\edth}\lambda_1+\frac{1}{2}v^e\partial_e\lambda
 _0+\Bigl(\rho'-\frac{1}{2}v^e\bigl({\cal D}_e\iota_A\bigr)o^A\Bigr)
 \lambda_0+\frac{1}{2}v^e\bigl({\cal D}_eo_A\bigr)o^A\lambda_1,
 \label{eq:2.1.5b}
\end{eqnarray}
where ${\edth}$ and ${\edth}'$ are the standard edth operators and 
$\rho:=\iota^A\bar o^{A'}(\nabla_{AA'}o_B)o^B$ and $\rho':=-o^A\bar
\iota^{A'}(\nabla_{AA'}\iota_B)\iota^B$, the convergences of the 
outgoing and incoming future pointing null normals of the surfaces 
${\cal S}_r$ in spacetime, respectively. (For the details see e.g. 
\cite{PR,HT}). 

For later use, let us write down the GHP form of the two-dimensional 
Sen--Dirac operator $\Delta_{A'A}\lambda^A$ and the 2-surface 
twistor operator ${\cal T}_{A'AB}{}^C\lambda_C$. These are 
\begin{align}
&\bar o^{A'}\Delta_{A'A}\lambda^A=-\bigl({\edth}'\lambda_0+\rho
 \lambda_1\bigr), \hskip 50pt
\bar\iota^{A'}\Delta_{A'A}\lambda^A=\bigl({\edth}\lambda_1+\rho'
 \lambda_0\bigr), \label{eq:2.1.D} \\
&\bar o^{A'}\iota^A\iota^B{\cal T}_{A'AB}{}^C\lambda_C=\bigl({\edth}'
 \lambda_1+\sigma'\lambda_0\bigr), \hskip 20pt
\bar\iota^{A'}o^Ao^B{\cal T}_{A'AB}{}^C\lambda_C=\bigl({\edth}
 \lambda_0+\sigma\lambda_1\bigr), \label{eq:2.1.T}
\end{align}
where $\sigma:=o^A\bar\iota^{A'}(\nabla_{AA'}o_B)o^B$ and $\sigma':=
-\iota^A\bar o^{A'}(\nabla_{AA'}\iota_B)\iota^B$, the shears of the 
null normals of ${\cal S}_r$. The 2-surface twistor equation or 
the equations defining the holomorphic or anti-holomorphic spinor 
fields are appropriate direct sums of these `elementary' 
differential operators. 


\subsection{Witten spinors and geometric triads}
\label{sub-2.2}

In this subsection, we summarize the key results of \cite{J} on 
the relationship between the solutions of the Witten equation 
and vector bases on $\Sigma$ that are orthonormal up to an 
overall function. 

If $t_{AA'}$ is the (e.g. future pointing) unit normal of $\Sigma$, 
then $G_{AA'}:=\sqrt{2}t_{AA'}$ is a positive definite Hermitian 
metric on the spin spaces, by means of which the primed spinor 
indices can be converted to unprimed ones according to $\bar
\lambda^{A'}\mapsto\lambda^\dagger{}^A:=-G^A{}_{A'}\bar\lambda^{A'}$ 
and $\bar\lambda_{A'}\mapsto\lambda^\dagger_A:=G_A{}^{A'}\bar\lambda
_{A'}$. Then $\rho^2:=G_{AA'}\lambda^A\bar\lambda^{A'}$, the norm 
of any non-zero spinor field $\lambda^A$ on $\Sigma$, is non-zero 
and $\{\lambda^A,\lambda^\dagger{}^A\}$ form a basis in the spin 
spaces. Such a spinor determines a vector basis $\{X^a,Y^a,Z^a\}$ 
on $\Sigma$ by 
\begin{equation}
\frac{1}{\sqrt{2}}\bigl(X^a+{\rm i}Y^a\bigr):=\lambda^{(A}\lambda
^{B)}, \hskip 20pt
\frac{1}{\sqrt{2}}Z^a:=\lambda^\dagger{}^{(A}\lambda^{B)}.
\label{eq:2.2.1}
\end{equation}
(N.B.: The standard convention $a=AA'$ of \cite{PR,HT} for the 
Lorentzian tensor and $SL(2,\mathbb{C})$ spinor indices yields 
that a {\em spatial} tensor index is identified with a pair of 
{\em symmetric} unprimed $SU(2)$ spinor indices.) 
$\{X^a,Y^a,Z^a\}$ is a real orthogonal basis, the length of each of 
these vectors is $\rho^2$, and $\varepsilon_{abc}X^aY^bZ^c=\rho^6$, 
where $\varepsilon_{abc}$ is the induced volume 3-form on $\Sigma$. 
Rewriting the derivatives $\bar\lambda^{A'}{\cal D}_{A'A}\lambda^A$ 
and $\lambda^BG_B{}^{A'}{\cal D}_{A'A}\lambda^A$ in terms of these 
vectors and taking into account that on the domain $U\subset\Sigma$ 
where $\lambda^A$ is not vanishing $\{\lambda^A,\lambda^\dagger{}^A\}$ 
is a basis in the spin spaces, the Witten equation is equivalent to
\begin{align}
&D_aX^a=0, \hskip 20pt D_aY^a=0, \hskip 20pt D_aZ^a=-\chi\rho^2,
\label{eq:2.2.2a}\\
&Z^aY^bD_aX_b+\bigl(Y^aX^b-X^aY^b\bigr)D_aZ_b=0. \label{eq:2.2.2b}
\end{align}
Finally, introducing the orthonormal basis $(E^a_1,E^b_2,E^a_3):=
\rho^{-2}(X^a,Y^a,Z^a)$ on the domain $U$, and if $\{\vartheta
^{\bi}_a\}$, ${\bi}=1,2,3$, is the dual 1-form basis on $U$ and 
$\gamma^{\bi}{}_{\bk\bj}:=\vartheta^{\bi}_aE^b_{\bk}D_bE^a_{\bj}$, 
the Ricci rotation coefficients, then these conditions can be 
rewritten as 
\begin{equation}
\gamma^{\bi}{}_{\bi\bj}=-E^a_{\bj}D_a\ln\rho^2-\chi\delta^3_{\bj}, 
\hskip 20pt
\gamma_{\bi\bj\bk}\varepsilon^{\bi\bj\bk}=0. \label{eq:2.2.3}
\end{equation}
Here boldface indices are raised and lowered by the constant 
negative definite metric $\eta_{\bi\bj}:=-\delta_{\bi\bj}$, and 
$\varepsilon_{\bi\bj\bk}:=\varepsilon_{abc}E^a_{\bi}E^b_{\bj}E^c
_{\bk}$. The frame gauge condition suggested in \cite{Ne,N1,N2} 
is equivalent to
\begin{equation}
\gamma^{\bi}{}_{\bi\bj}=-E^a_{\bj}D_a\ln\rho^2, \hskip 20pt
\gamma_{\bi\bj\bk}\varepsilon^{\bi\bj\bk}=-m
\label{eq:2.2.4}
\end{equation}
for some constant $m$. For asymptotically Cartesian frames on 
asymptotically flat 3-geometries $(\Sigma,h_{ab})$ this constant 
is vanishing, but it has a non-zero value for frames e.g. on 
$\Sigma\approx S^3$ \cite{Ne,N1,N2}. Thus, for $m=0$, the frame 
gauge condition of \cite{Ne,N1,N2} and the Witten gauge condition 
are the same on {\em maximal} hypersurfaces, and hence the present 
investigations have relevance from the point of view of both.


\subsection{The boundary conditions}
\label{sub-2.3}

\subsubsection{The general linear, first order boundary conditions}
\label{sub-2.3.1}

In typical problems of general relativity the hypersurface $\Sigma$ 
is either asymptotically flat/hyperboloidal with or without inner 
boundary, or compact with outer boundary. In the former case the 
solution to the Witten equation is usually assumed to be either 
asymptotically constant or a solution to the asymptotic twistor 
equation, depending on whether $\Sigma$ extends to spatial or null 
infinity. The inner boundary, and, in the latter case, the outer 
boundary will be denoted by ${\cal S}$, which is assumed to be a 
(not necessarily connected) closed orientable spacelike surface. 

Since the Witten equation is elliptic, only `half' of the data may 
be specified on ${\cal S}$ (see e.g. \cite{D} and references 
therein). In particular, the most general $\mathbb{C}$-linear, first 
order boundary condition for the Witten equation ${\cal D}_{A'A}
\lambda^A=0$ is 
\begin{equation}
f=\lambda^AA_A+v^e\bigl({\cal D}_e\lambda^A\bigr)v_{AA'}\bar B
^{A'}, \label{eq:2.3.1}
\end{equation}
where $f$ is a given function, while $A_A$ and $B^A$ are given 
spinor fields on the boundary 2-surface ${\cal S}$. This is a mixed, 
inhomogeneous boundary condition, which for vanishing $A_A$ is a 
purely Neumann, while for vanishing $B^A$ is a purely Dirichlet type 
boundary condition. The boundary condition is called homogeneous if 
$f=0$. The spinor field $\gamma^A{}_B$, introduced in the previous 
subsection, defines a chirality on the spinor bundle over ${\cal S}$, 
and the basis vectors $\iota^A$ and $o^A$ of a GHP spin frame are 
right handed/left handed spinors with respect to this. (For the 
details of this notion of chirality, see \cite{Sz1}.) The boundary 
condition is called {\em chiral} if the spinor fields $A_A$ and 
$B^A$ are proportional to the right or left handed eigenspinors, 
$\iota^A$ or $o^A$, of the chirality operator on ${\cal S}$. 

The advantage of the extra normal vector field $v_{AA'}$ in equation 
(\ref{eq:2.3.1}) (contracted with $\bar B^{A'}$) is that it makes 
the boundary condition 2+2 covariant. Indeed, using the decomposition 
(\ref{eq:2.1.3}) of the Witten equation at the points of the boundary 
${\cal S}$, the boundary condition (\ref{eq:2.3.1}) can be rewritten 
as 
\begin{equation}
f=\lambda^AA_A+\bar B^{A'}\bigl(\Delta_{A'A}\lambda^A\bigr),
\label{eq:2.3.2}
\end{equation}
which is manifestly 2+2 covariant. Equation (\ref{eq:2.1.3}) implies, 
in particular, that a Neumann type boundary condition for the Witten 
equation can always be written as a condition on the derivatives of 
the spinor field in the directions {\em tangential} to ${\cal S}$. 
A more general (only ${\mathbb R}$- rather than ${\mathbb
C}$-linear) boundary condition is 
\begin{equation}
f=\lambda^AA_A+\bar B^{A'}\bigl(\Delta_{A'A}\lambda^A\bigr)+\bar
E^{A'}\bar\gamma_{A'B'}\bar\lambda^{B'}
\label{eq:2.3.3}
\end{equation}
for some given function $f$ and spinor fields $A_A$, $B^A$ and 
$E^A$ on ${\cal S}$.

\subsubsection{Conformally invariant boundary conditions}
\label{sec-2.3.2}

Next let us clarify how the general boundary condition (\ref{eq:2.3.3}) 
changes under a conformal rescaling of the spacetime metric. Suppose 
that $\lambda_A$ has conformal weight $w$. Then, using $\Pi^a_b=\frac{1}
{2}(\delta^A_B\delta^{A'}_{B'}-\gamma^A{}_B\bar\gamma^{A'}{}_{B'})$ and 
the fact that $\gamma^A{}_B$ is trace free, it is easy to derive how 
$\Delta_{A'A}\lambda^A$ transforms under spacetime conformal rescalings. 
We obtain $\hat\Delta_{A'A}\hat\lambda^A=\Omega^{w-1}(\Delta_{A'A}
\lambda^A+\Upsilon_{A'A}\lambda^A+(w-1)\Upsilon_e\Pi^e_{A'A}\lambda^A)$.
Then multiplying equation (\ref{eq:2.3.3}) by $\Omega^{w-1}$ and 
expressing every field in terms of the conformally rescaled ones, we 
obtain 
\begin{align}
\Omega^{w-1}f&=\hat\lambda^A\Bigl(A_A-\Upsilon_{AA'}\bar B^{A'}-
 (w-1)\Upsilon_e\Pi^e_{AA'}\bar B^{A'}\Bigr)+\bar B^{A'}\bigl(\hat
 \Delta_{A'A}\hat\lambda^A\bigr)+\nonumber \\
&+\Omega^{-1}\bar E^{A'}\hat{\bar\gamma}_{A'B'}\hat{\bar\lambda}{}^{B'}.
 \label{eq:2.3.4}
\end{align}
Thus, the boundary condition (\ref{eq:2.3.3}) for the conformal 
weight $w$ spinor field $\lambda_A$ is conformally {\em covariant} 
if $B^A\mapsto\hat B^A:=B^A$, $A_A\mapsto\hat A_A:=A_A-\Upsilon_{AA'}
\bar B^{A'}-(w-1)\Upsilon_e\Pi^e_{AA'}\bar B^{A'}$, $E^A\mapsto\hat E
^A:=\Omega^{-1}E^A$ and $f\mapsto\hat f:=\Omega^{w-1}f$; i.e. in 
particular, $B^A$ has conformal weight zero, $E^A$ has conformal 
weight $-1$, and $f$ has conformal weight $(w-1)$. 
For $w=1$ and $E^A=0$ these are precisely the defining transformation 
properties of local twistors \cite{PR,HT}, in which case $f$ is just 
${\rm i}$ times the conformally {\em invariant} Hermitian scalar 
product of the local twistors $(\lambda^A,{\rm i}\Delta_{A'B}\lambda
^B)$ and $(B^A,-{\rm i}\bar A_{A'})$. 
For $w\not=1$, the boundary condition is conformally {\em invariant\/} 
if we impose the additional restriction $f=0$. Next we determine how 
the spinor field $A_A$ must be built to have the required 
transformation property above. 

Since $B^A$ has zero conformal weight, ${\rm i}\Delta_{A'A}B^A$ 
transforms under conformal rescalings as the secondary part of a 
contravariant twistor, i.e. $\hat\Delta_{AA'}\hat{\bar B}{}^{A'}=
\Delta_{AA'}\bar B^{A'}+\Upsilon_{AA'}\bar B^{A'}$. Thus, 
\begin{equation}
\hat A_A+\hat\Delta_{AA'}\hat{\bar B}{}^{A'}=A_A+\Delta_{AA'}\bar B
^{A'}-(w-1)\Upsilon_e\Pi^e_{AA'}\bar B^{A'}. \nonumber
\end{equation}
The last term on the right-hand side can, however, be compensated 
by using the 
(trace of the) spinor form $Q^A{}_{EE'B}:=\frac{1}{2}(\Delta_{EE'}
\gamma^A{}_C)\gamma^C{}_B$ of the extrinsic curvature tensor of ${\cal 
S}$. ($Q^A{}_{EE'B}$ is in fact the spinor form of the extrinsic 
curvature tensor, because the GHP convergences and shears can also 
be given in terms of this spinor as $\rho=o^A\iota^E\bar o^{E'}o^B
Q_{AEE'B}$, $-\rho'=\iota^Ao^E\bar\iota^{E'}\iota^BQ_{AEE'B}$, $\sigma
=o^Ao^E\bar\iota^{E'}o^BQ_{AEE'B}$ and $-\sigma'=\iota^A\iota^E\bar o
^{E'}\iota^BQ_{AEE'B}$ \cite{Sz1}. The mean curvature vector of ${\cal 
S}$ in the spacetime is $H_{AA'}:=2Q^E{}_{EA'A}=-2\rho\iota_A\bar\iota
_{A'}-2\rho'o_A\bar o_{A'}$, which is real. Since $H^{AA'}H_{A'B}=4\rho
\rho'\delta^A_B$, it defines an isomorphism between the spin and 
complex conjugate spin spaces precisely when $\rho\rho'\not=0$.) 
Indeed, a simple calculation yields that under a conformal rescaling 
of the spacetime metric $Q^A{}_{AE'E}\mapsto\hat Q^A{}_{AE'E}:=Q^A{}
_{AE'E}+\Upsilon_f(\delta^f_e-\Pi^f_e)$, and hence 
\begin{equation}
\hat Q^E{}_{EA'A}\hat{\bar B}{}^{A'}=Q^E{}_{EA'A}\bar B^{A'}+\hat
\Delta_{AA'}\hat{\bar B}{}^{A'}-\Delta_{AA'}\bar B^{A'}-\Upsilon_e
\Pi^e_{AA'}\bar B^{A'}. \nonumber
\end{equation}
Expressing the last term from this equation and substituting into the 
previous one, we obtain that 
\begin{equation}
\hat A_A+w\hat\Delta_{AA'}\hat{\bar B}{}^{A'}-\bigl(w-1\bigr)\hat Q
^E{}_{EA'A}\hat{\bar B}{}^{A'}=A_A+w\Delta_{AA'}\bar B^{A'}-(w-1)Q^E
{}_{EA'A}\bar B^{A'}, \nonumber
\end{equation}
i.e. the expression on the right-hand side has zero conformal 
weight. Therefore, {\em for any spinor field $C_A$ with zero 
conformal weight, the spinor field 
\begin{equation}
A_A:=C_A-w\Delta_{AA'}\bar B^{A'}+(w-1)Q^E{}_{EA'A}\bar B^{A'}
\label{eq:2.3.5}
\end{equation}
has the desired conformal transformation property}. Consequently, the 
${\mathbb R}$-linear, conformally invariant first order boundary 
condition for the Witten equation must have the form 
\begin{equation}
\lambda^A\Bigl(C_A+\frac{1}{2}\Delta_{AA'}\bar B^{A'}-\frac{3}{2}Q^E
{}_{EA'A}\bar B^{A'}\Bigr)+\bar B^{A'}\bigl(\Delta_{A'A}\lambda^A
\bigr)+\bar E^{A'}\bar\gamma_{A'B'}\bar\lambda^{B'}=0. \label{eq:2.3.6}
\end{equation}
Here $C_A$ and $B^A$ are arbitrary spinor fields with zero conformal 
weight and $E^A$ is an arbitrary spinor field with conformal weight 
$-1$ on ${\cal S}$. 


\section{Explicit solutions}
\label{sec-3}

\subsection{A solution of the Witten equation with one dimensional
zero-sets}
\label{sub-3.1}

In \cite{bar}, B\"ar showed that the set of zeros of the solutions of 
a Dirac equation on an $n$ dimensional Riemannian manifold is of 
dimension not greater than $(n-2)$. In this subsection, we illustrate 
this by a simple solution of the Witten equation in flat 3-space. 
Its zeros form, in fact, a union of discrete zeros and one dimensional 
submanifolds, which may be compact or non-compact (in fact, not 
bounded); or the set of zeros can consist of purely isolated points. 

Suppose that $\Sigma$ is both intrinsically and extrinsically flat, 
i.e. it is a hyperplane in Minkowski spacetime. Let $x^{\bi}=(x,y,z)$, 
${\bi}=1,2,3$, be a Cartesian coordinate system on $\Sigma\approx
\mathbb{R}^3$, i.e. in which $h_{\bi\bj}=-\delta_{\bi\bj}$, let 
$\sigma^{\uA}_{{\bi}{\uB}}$ be the $SU(2)$ Pauli matrices (divided 
by $\sqrt{2}$), and introduce the notation $D^{\uA}{}_{\uB}:=h
^{\bi\bj}\sigma^{\uA}_{{\bi}{\uB}}\partial_{\bj}$. Then in the 
Cartesian spin frame adapted to the coordinates above the Witten 
equation, $D^{\uA}{}_{\uB}\lambda^{\uB}=0$, takes the explicit form 
\begin{equation}
\partial_z\lambda^0+\partial_x\lambda^1+{\rm i}\partial_y\lambda^1
=0, \hskip 20pt
\partial_z\lambda^1-\partial_x\lambda^0+{\rm i}\partial_y\lambda^0
=0. \label{eq:3.1.1}
\end{equation}
A particular solution of these equations is 
\begin{equation*}
\lambda^0=-\bigl(A+B\bigr)z^2+Ax^2+By^2-C, \hskip 20pt
\lambda^1=2z\bigl(Ax-{\rm i}By\bigr)-D,
\end{equation*}
where $A$, $B$, $C$ and $D$ are constants. 
If $D=0$, then for real $A$, $B$ and $C$ the coordinates of its zeros 
satisfy $z=0$ and $Ax^2+By^2=C$. If the parameters are all positive, 
then the set of zeros is an ellipse, which is a compact 
one-dimensional submanifold. 
However, 
for $AB<0$ and $C\not=0$ the set of zeros is a hyperbola, while for 
$C=0$ it is a pair of straight lines crossing each other at the 
origin and the coordinates of these zeros are given by $(x,\pm x
\sqrt{-A/B},0)$. These sets are not bounded in $\mathbb{R}^3$. If 
$B=0$ and $AC>0$, then $\lambda^A$ has two isolated zeros at $(\pm
\sqrt{C/A},0,0)$. The vector basis corresponding to the constant 
spinor field $(\lambda^0,\lambda^1)=(-C,0)$ via (\ref{eq:2.2.1}) is 
just the constant orthonormal triad $E^a_{\bi}=-(\frac{\partial}
{\partial x^{\bi}})^a$.


\subsection{Spherically symmetric solutions in flat 3-space}
\label{sub-3.2}

In the present subsection we determine the fundamental solution of 
the Witten equation and discuss various explicit boundary conditions 
on spherically symmetric ${\cal S}$ that specify, among others, the 
constant spinor fields. 

\subsubsection{The fundamental solution}
\label{sub-3.2.1}

Let ${\cal D}_e$ be the {\em intrinsically and extrinsically flat} 
Sen derivative operator. The surfaces ${\cal S}_r$ of the foliation 
of $\Sigma$ will be chosen to be metric spheres of radius $r$. By 
the vanishing of the extrinsic curvature (indeed by $\chi=0$), one 
has $\rho=-1/r$ and $\rho'=1/2r$. Though the GHP spin frame $\{o^A,
\iota^A\}$ is not constant on $\Sigma$ with respect to the flat 
${\cal D}_e$, its normal directional derivative is vanishing: $v^e
{\cal D}_eo_A=v^e{\cal D}_e\iota_A=0$. Substituting these into 
(\ref{eq:2.1.5a})-(\ref{eq:2.1.5b}) we obtain 
\begin{equation}
{\edth}'\lambda_0-\frac{1}{r}\lambda_1-v^e\partial_e\lambda_1
=0, \hskip 20pt
{\edth}\lambda_1+\frac{1}{2r}\lambda_0+\frac{1}{2}v^e\partial_e
\lambda_0=0. \label{eq:3.2.1}
\end{equation}
Recalling that the spinor components $\lambda_0$ and $\lambda_1$ 
have spin weight $\frac{1}{2}$ and $-\frac{1}{2}$, respectively, 
we can expand them in terms of the $\pm\frac{1}{2}$ spin weighted 
spherical harmonics according to 
\begin{equation}
\lambda_0=\sum_{j=\frac{1}{2}}^\infty\sum_{m=-j}^jc_0^{jm}\bigl(r
\bigr){}_{\frac{1}{2}}Y_{jm}, \hskip 20pt
\lambda_1=\sum_{j=\frac{1}{2}}^\infty\sum_{m=-j}^jc_1^{jm}\bigl(r
\bigr){}_{-\frac{1}{2}}Y_{jm}. \label{eq:3.2.2}
\end{equation}
Note that $j$ takes the values $\frac{1}{2}$, $\frac{3}{2}$,
$\frac{5}{2}$, \dots\ and, for given $j$, $m=-j$, $-j+1$, \dots, $j$. 

Recalling (see e.g. \cite{PR}) that the action of the ${\edth}$ 
and ${\edth}'$ operators on the spin $s$ weighted spherical 
harmonics ${}_sY_{jm}$ is 
\begin{equation}
{\edth}{}_sY_{jm}=-\frac{1}{\sqrt{2}r}\sqrt{\bigl(j+s+1\bigr)
\bigl(j-s\bigr)}{}_{s+1}Y_{jm}, \hskip 10pt
{\edth}'{}_sY_{jm}=\frac{1}{\sqrt{2}r}\sqrt{\bigl(j-s+1\bigr)
\bigl(j+s\bigr)}{}_{s-1}Y_{jm}, \label{eq:3.2.3}
\end{equation}
(\ref{eq:3.2.1}) reduces to the system of ordinary differential 
equations 
\begin{equation}
\frac{{\rm d}c_0^{jm}}{{\rm d}r}+\frac{1}{r}c_0^{jm}-\frac{\sqrt2}
{r}\bigl(j+\frac{1}{2}\bigr)c_1^{jm}=0, \hskip 20pt
\frac{{\rm d}c_1^{jm}}{{\rm d}r}+\frac{1}{r}c_1^{jm}-\frac{1}
{\sqrt{2}r}\bigl(j+\frac{1}{2}\bigr)c_0^{jm}=0. \label{eq:3.2.4}
\end{equation}
These equations yield the second order equation  
\begin{equation}
c''+\frac{3}{r}c'+\frac{1}{r^2}\bigl[1-\bigl(j+\frac{1}{2}\bigr)^2
\bigr]c=0 \label{eq:3.2.5}
\end{equation}
both for $c=c_0^{jm}$ and $c_1^{jm}$, where the prime denotes 
differentiation with respect to $r$. Multiplying this with $r^k$ 
for some real $k$, we obtain 
\begin{equation}
\bigl(r^kc\bigr)''+\frac{3-2k}{r}\bigl(r^kc\bigr)'+\frac{1}{r^2}
\bigl[k^2-2k+1-\bigl(j+\frac{1}{2}\bigr)^2\bigr]\bigl(r^kc\bigr)=0.
\label{eq:3.2.6}
\end{equation}
Thus, to simplify this equation, let us choose $k$ to make the last 
term vanish, i.e. let $k=1\pm(j+\frac{1}{2})$. Then (\ref{eq:3.2.6}) 
can be integrated directly: $c(r)=Cr^{k-2}$, i.e. 
\begin{equation}
c^{jm}\bigl(r\bigr)={}_\pm C^{jm}r^{-1\pm(j+\frac{1}{2})},
\label{eq:3.2.7}
\end{equation}
where ${}_\pm C^{jm}$ are constants. Substituting this both for 
$c^{jm}_0$ and $c^{jm}_1$ back into the first order equations 
(\ref{eq:3.2.4}), we find that the sign $\pm$ in the exponent in 
(\ref{eq:3.2.7}) for $c^{jm}_0$ and $c^{jm}_1$ is the same; 
furthermore $\sqrt{2}{}_\pm C_1^{jm}=\pm{}_\pm C_0^{jm}$. 
Therefore, in the solution only the coefficients ${}_\pm C_0^{jm}$ 
will appear, and in the rest of this paper we use the notation $A
^{jm}:={}_+C_0^{jm}$, $B^{jm}:={}_-C_0^{jm}$. Thus, 
\begin{eqnarray}
\lambda_0\!\!\!\!&=\!\!\!\!&\sum_{j,m}A^{jm}r^{-1+(j+\frac{1}{2})}
 {}_{\frac{1}{2}}Y_{jm}+\sum_{j,m}B^{jm}r^{-1-(j+\frac{1}{2})}
 {}_{\frac{1}{2}}Y_{jm}, \label{eq:3.2.8a} \\
\sqrt{2}\lambda_1\!\!\!\!&=\!\!\!\!&\sum_{j,m}A^{jm}r^{-1+(j+
 \frac{1}{2})}{}_{-\frac{1}{2}}Y_{jm}-\sum_{j,m}B^{jm}r^{-1-(j+
 \frac{1}{2})}{}_{-\frac{1}{2}}Y_{jm}. \label{eq:3.2.8b}
\end{eqnarray}
This solution is analogous to the fundamental solution of the flat 
space Laplace equation with centre $r=0$, thus we may call it the 
fundamental solution of the Witten equation. Its general solution 
is a superposition of such fundamental solutions with different 
centres, and it is the boundary conditions that specify the actual 
solutions that we are interested in. 

For example, if $\Sigma$ is isometric with $\mathbb{R}^3$ and we 
want that the solutions be bounded at infinity, then by 
(\ref{eq:3.2.8a})--(\ref{eq:3.2.8b}) all the constants $A^{jm}$ must 
be zero for $j\geq\frac{3}{2}$. Then $\lambda_A$ tends to a constant 
spinor field at infinity, and the asymptotic value of $\lambda_A$ is 
fixed by $A^{\frac{1}{2}\pm\frac{1}{2}}$. Hence, the form of the spinor 
components is given by 
\begin{align}
\lambda_0&=\sum_{m=-\frac{1}{2}}^{\frac{1}{2}}\bigl(A^{\frac{1}{2}m}
 +\frac{B^{\frac{1}{2}m}}{r^2}\bigr){}_{\frac{1}{2}}Y_{\frac{1}{2}m}
 +\sum^\infty_{j=\frac{3}{2}}\sum^j_{m=-j}\frac{B^{jm}}{r^{1+(j+
 \frac{1}{2})}}{}_{\frac{1}{2}}Y_{jm}, \label{eq:3.2.9a} \\
\sqrt{2}\lambda_1&=\sum_{m=-\frac{1}{2}}^{\frac{1}{2}}\bigl(
 A^{\frac{1}{2}m}-\frac{B^{\frac{1}{2}m}}{r^2}\bigr){}_{-\frac{1}{2}}
 Y_{\frac{1}{2}m}-\sum^\infty_{j=\frac{3}{2}}\sum^j_{m=-j}\frac{
 B^{jm}}{r^{1+(j+\frac{1}{2})}}{}_{-\frac{1}{2}}Y_{jm}.
 \label{eq:3.2.9b}
\end{align}
On the other hand, these can be regular at the origin precisely when 
$B^{jm}=0$ for all $j$, in which case $\lambda_A$ is constant on 
$\Sigma$. There are two such linearly independent spinor fields, which 
are parametrized by the constants $A^{\frac{1}{2}\pm\frac{1}{2}}$. 
Similarly, if $\Sigma$ is isometric to ${\mathbb R}^3-\{0\}$, then 
the solution is bounded precisely when it is constant.

\subsubsection{Boundary conditions: the compact case}
\label{sub-3.2.2}

Suppose that $\Sigma$ is isometric to the solid ball $B\subset
{\mathbb R}^3$ of radius $R$, i.e. $\Sigma$ is compact with boundary 
$\partial\Sigma={\cal S}_R$. Then, to ensure the regularity of the 
spinor field (\ref{eq:3.2.8a})--(\ref{eq:3.2.8b}) at $r=0$, all the 
coefficients $B^{jm}$ must be zero, and hence 
\begin{equation}
\lambda_0=\sum_{j=\frac{1}{2}}^\infty\sum_{m=-j}^jA^{jm}r^{j-
   \frac{1}{2}}{}_{\frac{1}{2}}Y_{jm}, \hskip 15pt
\sqrt{2}\lambda_1=\sum_{j=\frac{1}{2}}^\infty\sum_{m=-j}^jA^{jm}
   r^{j-\frac{1}{2}}{}_{-\frac{1}{2}}Y_{jm}. \label{eq:3.2.10}
\end{equation}
This spinor field is completely fixed by one of the freely 
specifiable spinor components $\lambda_0$, $\lambda_1$, $v^e
\partial_e\lambda_0$ or $v^e\partial_e\lambda_1$, or at least by 
a combination of them, on ${\cal S}_R$. Next we discuss a few special 
cases. 

First, since the spin weighted spherical harmonics ${}_sY_{jm}$ form 
a basis in the space of the spin $s$ weighted functions on ${\cal S}
_R$, any homogeneous, chiral Dirichlet type boundary condition, e.g. 
$\lambda_1\vert_{{\cal S}_R}=0$, yields an identically zero spinor 
field. 

Next, it is easy to see that the inhomogeneous chiral boundary 
condition $\lambda_1\vert_{{\cal S}_R}=c^{\frac{1}{2}}{}_{-\frac{1}
{2}}Y_{\frac{1}{2}\frac{1}{2}}+c^{-\frac{1}{2}}{}_{-\frac{1}{2}}Y
_{\frac{1}{2}-\frac{1}{2}}$ yields a {\em constant} spinor field on 
$\Sigma$, where $c^{\pm\frac{1}{2}}$ are complex constants. Such 
spinor fields can also be characterized e.g. by ${\edth}'\lambda_1
\vert_{{\cal S}_R}=0$ (though the actual constant spinor field is not 
specified explicitly by this condition). This equation is just half 
of the equations defining the {\em holomorphic} spinor fields 
\cite{DM}, which equation appears in the 2-surface twistor equation 
too (see the first expression in (\ref{eq:2.1.T})). 
The constant spinor fields on $\Sigma$ can also be characterized on 
${\cal S}_R$ by conditions on the other component of the spinor 
field, e.g. by $\lambda_0\vert_{{\cal S}_R}=d^{\frac{1}{2}}{}
_{\frac{1}{2}}Y_{\frac{1}{2}\frac{1}{2}}+d^{-\frac{1}{2}}{}_{\frac{1}
{2}}Y_{\frac{1}{2}-\frac{1}{2}}$, where $d^{\pm\frac{1}{2}}$ are 
complex constants. This $\lambda_0$ on ${\cal S}_R$ is just the 
general solution of ${\edth}\lambda_0\vert_{{\cal S}_R}=0$, which is 
half of the equations defining the {\em anti-holomorphic} spinor 
fields on ${\cal S}_R$, as well as a half of the 2-surface twistor 
equation (see the second expression in (\ref{eq:2.1.T})). 

On the other hand, by (\ref{eq:3.2.10}) for fixed $j\geq\frac{3}{2}$ 
the inhomogeneous chiral boundary condition $\lambda_1\vert_{{\cal
S}_R}=\sum_{m=-j}^jc^m{}_{-\frac{1}{2}}Y_{jm}$ yields a non-constant 
spinor field, which {\em vanishes at the origin as $r^{j-\frac{1}
{2}}$}. 

Instead of chiral homogeneous boundary conditions involving only 
$\lambda_0$ or $\lambda_1$, we can consider the more general condition 
$(\edth'\lambda_0+\rho\lambda_1)\vert_{{\cal S}_R}=0$, where actually 
$\rho=-1/R$. Substituting (\ref{eq:3.2.10}) here, we obtain that $A
^{jm}=0$ for all $j\geq\frac{3}{2}$. Thus, only $A^{\frac{1}{2}\pm
\frac{1}{2}}$ may be nonzero, and hence the solution that this 
boundary condition singles out is constant. This boundary condition 
is just one-half of the defining equation of the holomorphic spinor 
fields on ${\cal S}_R$, and can also be written as $\bar o^{A'}\Delta
_{A'A}\lambda^A=0$ (see (\ref{eq:2.1.D})). Therefore, by the general 
considerations of section \ref{sub-2.3.1} (in particular by 
(\ref{eq:2.3.2})), or more explicitly by (\ref{eq:3.2.1}), this is 
just the chiral, homogeneous Neumann boundary condition $v^e({\cal D}
_e\lambda^A)\iota_A=0$. 
Similarly, we can impose $(\edth\lambda_1+\rho'\lambda_0)\vert_{{\cal
S}_R}=0$, where $\rho'=1/2R$. This specifies the constant solutions 
too, which boundary condition is just one-half of the defining 
equation of the anti-holomorphic spinor fields on ${\cal S}_R$. This 
can also be written as $\bar\iota^{A'}\Delta_{A'A}\lambda^A=0$ (see 
(\ref{eq:2.1.D})), or, equivalently, as the chiral, homogeneous 
Neumann boundary condition $v^e({\cal D}_e\lambda^A)o_A=0$. 

Thus there are several mathematically inequivalent ways to single 
out the constant spinor fields, but all these boundary conditions 
can be considered as weakening of the conditions defining the 
constant spinor fields {\em on the boundary} ${\cal S}_R$. (For 
further possibilities, see \cite{Sz2} and the Appendix of 
\cite{Sz3}.)

\subsubsection{Boundary conditions: the non-compact case}
\label{sub-3.2.3}

Now suppose that $\Sigma$ is isometric to $\overline{{\mathbb R}^3-
B}$, where $B\subset{\mathbb R}^3$ is the solid ball of radius $R>0$ 
and overline denotes topological closure in $\mathbb{R}^3$. 
Then $r$ is defined for $[R,\infty)$, and the inner boundary is 
${\cal S}_R$. By (\ref{eq:3.2.9a})--(\ref{eq:3.2.9b}) for given 
boundary conditions at infinity yielding asymptotically constant 
spinor fields (i.e. for fixed $A^{\frac{1}{2}\pm\frac{1}{2}}$), the 
solution $\lambda_A$ is completely determined by the coefficients 
$B^{jm}$. In particular, $\lambda_A$ is fixed by one of the freely 
specifiable spinor components $\lambda_0$, $\lambda_1$, $v^e\partial
_e\lambda_0$ or $v^e\partial_e\lambda_1$, or at least by specifying 
a combination of them, on ${\cal S}_R$. Next we discuss some 
particular cases. 

First consider $\lambda_0\vert_{{\cal S}_R}=0$, which is a homogeneous 
chiral Dirichlet type boundary condition. This was used in the proof 
of the positivity of the total (ADM and Bondi--Sachs) energy in the 
presence of black holes \cite{GH,RT}. Then, by the completeness of the 
spherical harmonics ${}_sY_{jm}$ in the space of the functions with 
spin weight $s$ on ${\cal S}_R$, (\ref{eq:3.2.9a}) implies that 
$B^{jm}=0$ for $j=\frac{3}{2}, \frac{5}{2}, \frac{7}{2}, \dots $, and 
that $B^{\frac{1}{2}m}=-R^2A^{\frac{1}{2}m}$. Therefore, 
\begin{equation}
\lambda_0=\bigl(1-\frac{R^2}{r^2}\bigr)\sum^{\frac{1}{2}}_{m=-
\frac{1}{2}}A^{\frac{1}{2}m}{}_{\frac{1}{2}}Y_{\frac{1}{2}m},
\hskip 20pt
\sqrt{2}\lambda_1=\bigl(1+\frac{R^2}{r^2}\bigr)\sum^{\frac{1}{2}}_{m
=-\frac{1}{2}}A^{\frac{1}{2}m}{}_{-\frac{1}{2}}Y_{\frac{1}{2}m},
\label{eq:3.2.11}
\end{equation}
which is a uniquely determined {\em non-constant} solution. The 
solution space is two dimensional, and can be coordinatized by 
$A^{\frac{1}{2}\pm\frac{1}{2}}$, or, equivalently, by the components 
of the spinor field at infinity. Recalling that the $s=\pm\frac{1}{2}$ 
spin weighted spherical harmonics for $j=\frac{1}{2}$ in the standard 
complex stereographic coordinates $\zeta:=\exp({\rm i}\phi)\cot
\frac{\theta}{2}$ take the form 
\begin{eqnarray}
{}_{\frac{1}{2}}Y_{\frac{1}{2}\frac{1}{2}}\!\!\!\!&=\!\!\!\!&
 \frac{\rm i}{\sqrt{2\pi}}\frac{\zeta}{\sqrt{1+\zeta\bar\zeta}},
 \hskip 20pt
 {}_{\frac{1}{2}}Y_{\frac{1}{2}-\frac{1}{2}}=\frac{\rm i}
 {\sqrt{2\pi}}\frac{1}{\sqrt{1+\zeta\bar\zeta}}, \label{eq:3.s.a} \\
{}_{-\frac{1}{2}}Y_{\frac{1}{2}\frac{1}{2}}\!\!\!\!&=\!\!\!\!&
 \frac{\rm i}{\sqrt{2\pi}}\frac{1}{\sqrt{1+\zeta\bar\zeta}},
 \hskip 14pt
 {}_{-\frac{1}{2}}Y_{\frac{1}{2}-\frac{1}{2}}=-\frac{\rm i}
 {\sqrt{2\pi}}\frac{\bar\zeta}{\sqrt{1+\zeta\bar\zeta}},
 \label{eq:3.s.b}
\end{eqnarray}
by $\lambda_0\vert_{{\cal S}_R}=0$ the {\em spinor field $\lambda_A$ 
has a zero on ${\cal S}_R$}. Indeed, if we write $\lambda_1\vert
_{{\cal S}_R}=a{}_{-\frac{1}{2}}Y_{\frac{1}{2}\frac{1}{2}}+b{}_{-
\frac{1}{2}}Y_{\frac{1}{2}-\frac{1}{2}}$, then $\lambda_1\vert
_{{\cal S}_R}$ is vanishing at $\zeta=\bar a/\bar b$ for non-zero 
$b$, while for $b=0$ it is vanishing at the `north pole' $\zeta=
\infty$ of ${\cal S}_R$. 

Instead of specifying $\lambda_0$ we can prescribe only its {\em 
tangential} derivatives. By (\ref{eq:3.2.9a}) 
\begin{equation}
{\edth}'\lambda_0=\frac{1}{\sqrt{2}r}\Bigl(\sum_{m=-\frac{1}{2}}
^{\frac{1}{2}}\bigl(A^{\frac{1}{2}m}+\frac{1}{r^2}B^{\frac{1}{2}m}
\bigr){}_{-\frac{1}{2}}Y_{\frac{1}{2}m}+\sum_{j=
\frac{3}{2}}^\infty\sum_{m=-j}^j(j+\frac{1}{2})\frac{B^{jm}}{r^{j+
\frac{3}{2}}}{}_{-\frac{1}{2}}Y_{jm}\Bigr), \label{eq:3.2.12}
\end{equation}
and hence if ${\edth}'\lambda_0\vert_{{\cal S}_R}=0$, then $B^{jm}
=0$ for $j=\frac{3}{2}, \frac{5}{2}, \dots $ and $B^{\frac{1}{2}m}=
-R^2A^{\frac{1}{2}m}$. Substituting these into 
(\ref{eq:3.2.9a})--(\ref{eq:3.2.9b}) we obtain (\ref{eq:3.2.11}) 
above, i.e. in particular, $\lambda_0\vert_{{\cal S}_R}=0$. Indeed, 
general theorems (see e.g. \cite{PR}) on the dimension of the kernel 
of the ${\edth}$ operators guarantee that ${\edth}'\lambda_0=0$ 
implies the vanishing of $\lambda_0$ itself on ${\cal S}_R$. 

Again by (\ref{eq:3.2.9a})
\begin{equation}
{\edth}\lambda_0=-\frac{1}{\sqrt{2}r}\sum_{j=\frac{3}{2}}^\infty
\sum_{m=-j}^j\sqrt{(j+\frac{3}{2})(j-\frac{1}{2})}\frac{B^{jm}}
{r^{j+\frac{3}{2}}}{}_{\frac{3}{2}}Y_{jm}. \label{eq:3.2.13}
\end{equation}
Thus, if our boundary condition is ${\edth}\lambda_0\vert_{{\cal S}
_R}=0$, which is one-half of the 2-surface twistor equation (see the 
second expression in (\ref{eq:2.1.T})), then $B^{jm}=0$ for $j=
\frac{3}{2},\frac{5}{2}, \dots\ $. Hence, 
\begin{equation}
\lambda_0=\sum^{\frac{1}{2}}_{m=-\frac{1}{2}}\bigl(A^{\frac{1}{2}m}+
\frac{1}{r^2}B^{\frac{1}{2}m}\bigr){}_{\frac{1}{2}}Y_{\frac{1}{2}m},
\hskip 20pt
\sqrt{2}\lambda_1=\sum^{\frac{1}{2}}_{m=-\frac{1}{2}}\bigl(A^{\frac{1}
{2}m}-\frac{1}{r^2}B^{\frac{1}{2}m}\bigr){}_{-\frac{1}{2}}Y_{\frac{1}
{2}m}. \label{eq:3.2.14}
\end{equation}
This boundary condition on ${\cal S}_R$ is equivalent to an {\em 
inhomogeneous} chiral Dirichlet type boundary condition of the 
form $\lambda_0\vert_{{\cal S}_R}=\sum_{m=-\frac{1}{2}}^{\frac{1}{2}}
c^m{}_{\frac{1}{2}}Y_{\frac{1}{2}m}$ with complex constants $c^m$. 
The space of the spinor fields (\ref{eq:3.2.14}) is {\em four} 
dimensional. Clearly, the investigation of the boundary condition 
${\edth}^n\lambda_0\vert_{{\cal S}_R}=0$ for any given $n\in{\mathbb
N}$ can be carried out similarly. For example, ${\edth}^2\lambda_0
\vert_{{\cal S}_R}=0$ is equivalent to an inhomogeneous one on 
$\lambda_0$, and yields {\em eight} complex dimensional space of 
solutions of the Witten equation. 

Finally, instead of equations for $\lambda_0$ {\em or} $\lambda_1$ 
on ${\cal S}_R$ we can impose the boundary condition $({\edth}'
\lambda_0+\rho\lambda_1)\vert_{{\cal S}_R}=0$. Then by 
(\ref{eq:3.2.9a})--(\ref{eq:3.2.9b}) this implies that $B^{jm}=0$ 
for all $j$, and hence the spinor field $\lambda_A$ is {\em constant 
on $\Sigma$}. Similarly, $({\edth}\lambda_1+\rho'\lambda_0)\vert
_{{\cal S}_R}=0$ also yields the constant spinor fields. As in the 
compact case, these are equivalent to chiral, homogeneous Neumann 
boundary conditions.


\subsection{Solutions on maximal, intrinsically conformally flat
hypersurfaces}
\label{sub-3.3}

Suppose that $\Sigma$ is maximal (i.e. $\chi=0$), the intrinsic 
metric $h_{ab}$ is conformally flat, and related to a flat metric 
$\hat h_{ab}$ by a {\em globally defined} conformal factor: $h_{ab}
=\Omega^{-2}\hat h_{ab}$. The $t={\rm const}$ hypersurfaces in the 
Reissner--Nordstr\"om spacetime, or the Brill--Lindquist \cite{BL} 
and Bowen--York data sets \cite{BoY} are such hypersurfaces. Here 
we solve the Witten equation on these data sets.

\subsubsection{Asymptotically constant solutions on complete
Reissner--Nordstr\"om data sets}
\label{sub-3.3.1}

The base manifold of the (maximally extended) Reissner--Nordstr\"om 
data set is $\Sigma\approx{\mathbb R}^3-\{0\}$ with the standard 
Cartesian coordinates $\{x^{\bi}\}$, ${\bi}=1,2,3$, or the 
corresponding spherical polar coordinates $(r,\theta,\phi)$. The 
conformal factor is given by $\Omega^{-1}(r):=(1+\frac{m}{2r})^2-
(\frac{e}{2r})^2$, where $m>\vert e\vert$. The extrinsic curvature 
of $\Sigma$ in the spacetime vanishes. The surface $r=\frac{1}{2}
\sqrt{m^2-e^2}$ is a stable minimal surface in $(\Sigma,h_{ab})$, 
whose points are just the fixed points of the discrete isometry 
$I:x^{\bi}\mapsto\frac{1}{4}(m^2-e^2)\frac{x^{\bi}}{r^2}$ (see 
\cite{Gi}). This surface represents the black hole event horizon in 
$\Sigma$, while the $r\rightarrow\infty$ and $r\rightarrow0$ regimes 
are the two asymptotically flat ends. 

Let $\hat\lambda^A$ be a solution of the flat Witten equation on 
$(\Sigma,\hat h_{ab})$. Then, by equation (\ref{eq:2.1.2}), 
$\lambda_A=\Omega^{\frac{1}{2}}\hat\lambda_A$ is a solution of the 
Witten equation on $(\Sigma,h_{ab})$. The conformal rescaling 
$\varepsilon_{AB}=\Omega^{-1}\hat\varepsilon_{AB}$ implies the 
rescaling $o^A=\Omega^{1-k}\hat o^A$, $\iota^A=\Omega^k\hat\iota
^A$ of the normalized spin frame with undetermined $k\in{\mathbb
R}$. We choose $k=\frac{1}{2}$ (the symmetric rescaling), so that 
\begin{equation*}
\lambda_Ao^A=\Omega\hat\lambda_A\hat o^A, \hskip 20pt
\lambda_A\iota^A=\Omega\hat\lambda_A\hat\iota^A.
\end{equation*}
Since $\Omega\rightarrow1$ if $r\rightarrow\infty$, the spinor field 
$\lambda_A$ can be non-singular on $\Sigma$ and bounded in this limit 
only if the components of $\hat\lambda_A$ in the spin frame $\{\hat o
^A,\hat\iota^A\}$ are given by (\ref{eq:3.2.9a})--(\ref{eq:3.2.9b}). 
However, since in the $r\rightarrow0$ limit $\Omega$ tends to zero as 
$r^2$, the solution $\lambda_A$ is bounded on the other asymptotic 
end precisely when $B^{jm}=0$ for all $j\geq\frac{3}{2}$, so that 
the spinor field is asymptotically constant there too. It is given 
explicitly by 
\begin{eqnarray}
\lambda_Ao^A\!\!\!\!&=\!\!\!\!&\frac{4r^2}{(2r+m)^2- e^2}\sum_{m=-
 \frac{1}{2}}^{\frac{1}{2}}\bigl(A^{\frac{1}{2}m}+\frac{B^{\frac{1}
 {2}m}}{r^2}\bigr){}_{\frac{1}{2}}Y_{\frac{1}{2}m}, 
 \label{eq:3.3.1.a} \\
\sqrt{2}\lambda_A\iota^A\!\!\!\!&=\!\!\!\!&\frac{4r^2}{(2r+m)^2-e^2}
 \sum_{m=-\frac{1}{2}}^{\frac{1}{2}}\bigl(A^{\frac{1}{2}m}-\frac{B
 ^{\frac{1}{2}m}}{r^2}\bigr){}_{-\frac{1}{2}}Y_{\frac{1}{2}m}.
 \label{eq:3.3.1.b}
\end{eqnarray}
Thus its asymptotic values are determined by $A^{\frac{1}{2}m}$ and 
$B^{\frac{1}{2}m}$. This solution can be considered as the sum of 
two spinor fields, one with $A^{\frac{1}{2}m}=0$ and the other with 
$B^{\frac{1}{2}m}=0$, and each is proportional to some constant 
spinor field with respect to some flat connection. However, their 
factors of proportionality are different, yielding different 
asymptotic properties: while one spinor field tends to a non-zero 
constant spinor at one asymptotic end and tends to zero at the other 
end, the other spinor field behaves in just the opposite way. 

By (\ref{eq:3.s.a})--(\ref{eq:3.s.b}), the spinor field $\lambda_A$ 
vanishes at the point $(r,\zeta,\bar\zeta)$ precisely when 
\begin{eqnarray}
\zeta\bigl(r^2A^{\frac{1}{2}\frac{1}{2}}+B^{\frac{1}{2}\frac{1}{2}}
 \bigr)+\bigl(r^2A^{\frac{1}{2}-\frac{1}{2}}+B^{\frac{1}{2}-\frac{1}
 {2}}\bigr)\!\!\!\!&=\!\!\!\!&0, \nonumber \\
\bigl(r^2A^{\frac{1}{2}\frac{1}{2}}-B^{\frac{1}{2}\frac{1}{2}}\bigr)
 -\overline{\zeta}\bigl(r^2A^{\frac{1}{2}-\frac{1}{2}}-B^{\frac{1}{2}
 -\frac{1}{2}}\bigr)\!\!\!\!&=\!\!\!\!&0. \nonumber
\end{eqnarray}
These equations yield 
\begin{equation*}
\zeta=-\frac{r^2A^{\frac{1}{2}-\frac{1}{2}}+B^{\frac{1}{2}-\frac{1}
{2}}}{r^2A^{\frac{1}{2}\frac{1}{2}}+B^{\frac{1}{2}\frac{1}{2}}}=
\frac{r^2\overline{A^{\frac{1}{2}\frac{1}{2}}}-\overline{B^{\frac{1}
{2}\frac{1}{2}}}}{r^2\overline{A^{\frac{1}{2}-\frac{1}{2}}}-\overline{
B^{\frac{1}{2}-\frac{1}{2}}}},
\end{equation*}
from which it follows that 
\begin{align}
&r^4\Bigl(\vert A^{\frac{1}{2}\frac{1}{2}}\vert^2+\vert A^{\frac{1}
 {2}-\frac{1}{2}}\vert^2\Bigr)- \nonumber \\
-&r^2\Bigl(A^{\frac{1}{2}\frac{1}{2}}\overline{B^{\frac{1}{2}
 \frac{1}{2}}}+A^{\frac{1}{2}-\frac{1}{2}}\overline{B^{\frac{1}{2}-
 \frac{1}{2}}}-\overline{A^{\frac{1}{2}\frac{1}{2}}}B^{\frac{1}{2}
 \frac{1}{2}}-\overline{A^{\frac{1}{2}-\frac{1}{2}}}B^{\frac{1}{2}-
 \frac{1}{2}}\Bigr)- \nonumber \\
&-\Bigl(\vert B^{\frac{1}{2}\frac{1}{2}}\vert^2+\vert B^{\frac{1}{2}
 -\frac{1}{2}}\vert^2\Bigr)=0. \label{eq:3.3.2}
\end{align}
However, the coefficient of $r^2$ is purely imaginary while all the 
other terms are real. Thus, to have a real solution for $r^2$, this 
imaginary coefficient must vanish, yielding one real condition for 
the coefficients $A^{\frac{1}{2}m}$ and $B^{\frac{1}{2}m}$. 
Parameterizing these coefficients as 
\begin{align*}
&A^{\frac{1}{2}\frac{1}{2}}=:\vert A\vert\cos\chi e^{{\rm i}\alpha
_1}, \hskip 20pt
A^{\frac{1}{2}-\frac{1}{2}}=:\vert A\vert\sin\chi e^{{\rm i}\alpha
_2}, \\
&B^{\frac{1}{2}\frac{1}{2}}=:\vert B\vert\cos\psi e^{{\rm i}\beta
_1}, \hskip 20pt
B^{\frac{1}{2}-\frac{1}{2}}=:\vert B\vert\sin\psi e^{{\rm i}\beta
_2},
\end{align*}
the condition that the coefficient of $r^2$ in (\ref{eq:3.3.2}) must 
be vanishing is 
\begin{equation}
\cos^2\psi=\frac{\sin^2\chi\sin^2\bigl(\alpha_2-\beta_2\bigr)}
{\cos^2\chi\sin^2\bigl(\alpha_1-\beta_1\bigr)+\sin^2\chi\sin^2
\bigl(\alpha_2-\beta_2\bigr)}. \label{eq:3.3.3}
\end{equation}
Then for the coordinates of the zero we obtain 
\begin{equation}
r^2=\frac{\vert B\vert}{\vert A\vert}, \hskip 20pt
\zeta=-\frac{\sin\chi e^{{\rm i}\alpha_2}+\sin\psi e^{{\rm i}
\beta_2}}{\cos\chi e^{{\rm i}\alpha_1}+\cos\psi e^{{\rm i}\beta_1}}.
\label{eq:3.3.4}
\end{equation}
Therefore, the solution $\lambda_A$ does not have any zero iff the 
parameters $A^{\frac{1}{2}m}$ and $B^{\frac{1}{2}m}$ do not satisfy 
(\ref{eq:3.3.3}). If (\ref{eq:3.3.3}) holds, then by the first of 
(\ref{eq:3.3.4}) the solution $\lambda_A$ has a zero for $r\in(0,
\infty)$ precisely when $A^{\frac{1}{2}m}\not=0$ and $B^{\frac{1}{2}
m}\not=0$, while the zero is at $r=\infty$ for $A^{\frac{1}{2}m}=0$ 
and at $r=0$ for $B^{\frac{1}{2}m}=0$. Moreover, even for fixed 
(nonzero) $A^{\frac{1}{2}m}$ and point $p=(r,\zeta,\bar\zeta)$ there 
is a parameter $B^{\frac{1}{2}m}$ such that the corresponding spinor 
field has a zero at $p$. These solutions are {\em not} conformal to 
a spinor field which would be constant with respect to some flat 
connection. Criterion (\ref{eq:3.3.3}) for the existence of the 
zeros determines a seven dimensional submanifold in the space of the 
parameters $(\vert A\vert,\chi,\alpha_1,\alpha_2;\vert B\vert,\psi,
\beta_1,\beta_2)$. Therefore, the solutions of the Witten equation 
on the non-extreme Reissner--Nordstr\"om data set that are constant 
at the two infinities form a four dimensional complex vector space, 
in which there is a seven dimensional real submanifold of solutions 
with a zero.

To interpret this result in the language of the geometric triads of 
subsection \ref{sub-2.2}, let us recall that a spinor determines an 
orthonormal triad only up to an overall real scale factor. Thus, the 
boundary conditions for the triads at the two asymptotic ends form 
a real 3+3 dimensional manifold (corresponding to the parameters 
$(\chi,\alpha_1,\alpha_2)$ and $(\psi,\beta_1,\beta_2)$ above). Since, 
however, the `modulus' of the spinor fields, $\vert A\vert$ and $\vert
B\vert$, are not involved in (\ref{eq:3.3.3}), it defines a five 
dimensional submanifold of boundary conditions for those frames which 
are singular somewhere inside the Reissner--Nordstr\"om initial data 
hypersurface. Thus, we do not have complete freedom (i.e. constant 
rotations) to choose the frame at the two infinities any way we want.

\subsubsection{Solutions on Reissner--Nordstr\"om data sets with
internal boundary}
\label{sub-3.3.2}

Let $\Sigma$ be the subset $\{\,x^{\bi}\,\vert\, \delta_{\bi\bj}x
^{\bi}x^{\bj}\geq R^2>0\,\}$ of the complete Reissner--Nordstr\"om 
data set of the previous subsection, whose inner boundary ${\cal S}$ 
is the 2-sphere with coordinate radius $R$. Consider the solutions 
$\lambda_A$ of the Witten equation that are asymptotically constant 
at the `outer' infinity $r\rightarrow\infty$. Then, by the discussion 
in the first half of the second paragraph of subsection 
\ref{sub-3.3.1}, the components of $\lambda_A$ are 
\begin{align}
\lambda_0&=\frac{4r^2}{(2r+m)^2-e^2}\Bigl(\sum_{m=-\frac{1}{2}}
 ^{\frac{1}{2}}\bigl(A^{\frac{1}{2}m}+\frac{B^{\frac{1}{2}m}}{r^2}
 \bigr){}_{\frac{1}{2}}Y_{\frac{1}{2}m}+\sum^\infty_{j=\frac{3}{2}}
 \sum^j_{m=-j}\frac{B^{jm}}{r^{1+(j+\frac{1}{2})}}{}_{\frac{1}{2}}
 Y_{jm}\Bigr), \label{eq:3.3.5a} \\
\sqrt{2}\lambda_1&=\frac{4r^2}{(2r+m)^2-e^2}\Bigl(\sum_{m=-\frac{1}
 {2}}^{\frac{1}{2}}\bigl(A^{\frac{1}{2}m}-\frac{B^{\frac{1}{2}m}}
 {r^2}\bigr){}_{-\frac{1}{2}}Y_{\frac{1}{2}m}-\sum^\infty_{j=\frac{3}
 {2}}\sum^j_{m=-j}\frac{B^{jm}}{r^{1+(j+\frac{1}{2})}}{}_{-\frac{1}
 {2}}Y_{jm}\Bigr). \label{eq:3.3.5b}
\end{align}
The areal radius of the spheres ${\cal S}_r$ of coordinate radius $r$, 
defined by $(\frac{1}{4\pi}{\rm Area}({\cal S}_r))^{\frac{1}{2}}$, is 
$r\Omega^{-1}=\frac{1}{4r}((2r+m)^2-e^2)$. Moreover, if the sign of 
the normal of ${\cal S}$ is chosen such that $\hat v^a=(\frac{\partial}
{\partial r})^a$, which points {\em inward on the hypersurface} 
$\Sigma$, then for the mean curvature of the 2-spheres ${\cal S}_r$ we 
obtain 
\begin{equation}
\nu=\frac{8r}{\bigl((2r+m)^2-e^2\bigr)^2}\Bigl(4r^2-\bigl(m^2-e^2
\bigr)\Bigr). \label{eq:3.3.6}
\end{equation}
Then, since the Reissner--Nordstr\"om data set is extrinsically flat, 
the outgoing and incoming null convergences on the 2-spheres ${\cal S}
_r$ are $\rho=-\frac{1}{2}\nu$ and $\rho'=\frac{1}{4}\nu$, 
respectively. Note that these are vanishing on the minimal surface 
$2r=\sqrt{m^2-e^2}$. Next we discuss a few explicit boundary 
conditions on ${\cal S}$. 

First, as in subsection \ref{sub-3.2.3}, any of the homogeneous, 
chiral Dirichlet boundary conditions, $\lambda_0\vert_{\cal S}=0$ or 
$\lambda_1\vert_{\cal S}=0$, specifies a solution on $\Sigma$ which 
is asymptotically constant at infinity but vanishing somewhere on 
${\cal S}$. This solution is conformal to that given by 
(\ref{eq:3.2.11}). Also as in subsection \ref{sub-3.2.3}, this 
solution can be characterized by the boundary condition ${\edth}'
\lambda_0=0$ or ${\edth}\lambda_1=0$, respectively. 

The purely right-handed or left-handed parts of the 2-surface twistor
equation, ${\edth}\lambda_0=0$ or ${\edth}'\lambda_1=0$, specify 
solutions conformal to the non-constant spinor fields (in the flat 
3-space) given by (\ref{eq:3.2.14}). 

However, the non-triviality of the conformal factor gives new 
possibilities. Clearly, 
\begin{align}
-\bar o^{A'}\Delta_{A'A}\lambda^A&={\edth}'\lambda_0+\rho\lambda_1=
 \Omega\Bigl(\Omega\hat{\edth}{}'\hat\lambda_0+\frac{1}{2}r\nu\hat
 \rho\hat\lambda_1\Bigr), \label{eq:3.3.7a} \\
\bar\iota^{A'}\Delta_{A'A}\lambda^A&={\edth}\lambda_1+\rho'\lambda
 _0=\Omega\Bigl(\Omega\hat{\edth}\hat\lambda_1+\frac{1}{2}r\nu\hat
 \rho'\hat\lambda_0\Bigr); \label{eq:3.3.7b}
\end{align}
where the hat refers to the flat 3-space: $\hat{\edth}$ is the flat 
space edth operator and $\hat\rho=-1/r$ and $\hat\rho'=1/2r$ are the 
flat spacetime convergences of subsection \ref{sub-3.2.3}, while the 
spinor components $\hat\lambda_0$ and $\hat\lambda_1$ are given by 
(\ref{eq:3.2.9a})--(\ref{eq:3.2.9b}). Then an apparently obvious 
choice for the boundary condition would be e.g. $\bar o^{A'}\Delta
_{A'A}\lambda^A=0$, as in subsection \ref{sub-3.2.3}. Then by the 
expression (\ref{eq:3.2.12}) for $\hat{\edth}{}'\hat\lambda_0$ this 
boundary condition yields $B^{jm}=0$ for all $j\geq\frac{3}{2}$ and 
$R^2A^{\frac{1}{2}m}+B^{\frac{1}{2}m}+N(R^2A^{\frac{1}{2}m}-B
^{\frac{1}{2}m})=0$, $m=\pm\frac{1}{2}$, where 
\begin{equation*}
N:=\frac{1}{2}\frac{4R^2-\bigl(m^2-e^2\bigr)}
{\bigl(2R^2+m\bigr)^2-e^2},
\end{equation*}
which is proportional to the mean curvature $\nu$ of the boundary 
(see equation (\ref{eq:3.3.6})), and takes its values between 
$-\frac{1}{2}$ and $\frac{1}{2}$. Thus, finally, the solution is 
given by 
\begin{equation}
\lambda_0=\Omega\Bigl(1-\frac{1+N}{1-N}\frac{R^2}{r^2}\Bigr)\sum
 _{m=-\frac{1}{2}}^{\frac{1}{2}}A^{\frac{1}{2}m}{}_{\frac{1}{2}}Y
 _{\frac{1}{2}m}, \hskip 15pt
\sqrt{2}\lambda_1=\Omega\Bigl(1+\frac{1+N}{1-N}\frac{R^2}{r^2}\Bigr)
 \sum_{m=-\frac{1}{2}}^{\frac{1}{2}}A^{\frac{1}{2}m}{}_{-\frac{1}{2}}
 Y_{\frac{1}{2}m} \label{eq:3.3.8b}
\end{equation}
for $r\geq R$. If $N>0$, i.e. if $\Sigma$ does not contain the 
minimal surface, then $\frac{1+N}{1-N}>1$, and hence there is a 
value $r=r_0>R$ for which $\lambda_0$ is vanishing. Then by 
(\ref{eq:3.s.a})--(\ref{eq:3.s.b}) the other spinor component is 
zero for some $(\zeta_0,\bar\zeta_0)$, and hence at the point $(r_0,
\zeta_0,\bar\zeta_0)$ the spinor field $\lambda_A$ is vanishing. If 
$N=0$, i.e. if the boundary ${\cal S}$ is just the minimal surface, 
then $\lambda_0=0$ there, and this case reduces to that of the 
homogeneous chiral Dirichlet boundary condition above. If $N<0$, i.e. 
when the minimal surface is contained in the interior of $\Sigma$, 
then the spinor field $\lambda_A$ does not have any zero. It might 
be worth noting that in the limit $R\rightarrow0$ (i.e. when the 
boundary ${\cal S}$ is `pushed out' to the other infinity) this 
spinor field tends to one of the two conformally constant spinor 
fields in (\ref{eq:3.3.1.a})--(\ref{eq:3.3.1.b}). 

Nevertheless, there is another `natural' choice for the boundary 
condition: $\hat{\edth}{}'\hat\lambda_0+\hat\rho\hat\lambda_1=0$, 
i.e. (by the results of subsection \ref{sub-3.2.3}) the condition 
that the spinor field $\hat\lambda_A$ be constant in the geometry of 
the flat 3-space. Then $\hat{\edth}{}\hat\lambda_1+\hat\rho'\hat
\lambda_0=0$ also holds, and hence a direct calculation yields that 
\begin{align}
-\bar o^{A'}\Delta_{A'A}\lambda^A&=\Omega\Bigl(-\Omega\hat\rho\hat
 \lambda_1+\frac{1}{2}r\nu\hat\rho\hat\lambda_1\Bigr)=-\frac{1}{r}
 \Bigl(\frac{1}{2}r\nu-\Omega\Bigr)\lambda_1=-\bar o^{A'}\Bigl(v^e
 \bigl(D_e\ln\Omega\bigr)v_{A'A}\lambda^A\Bigr) \label{eq:3.3.9a} \\
\bar\iota^{A'}\Delta_{A'A}\lambda^A&=\Omega\Bigl(-\Omega\hat\rho'\hat
 \lambda_0+\frac{1}{2}r\nu\hat\rho'\hat\lambda_0\Bigr)=\frac{1}{2r}
 \Bigl(\frac{1}{2}r\nu-\Omega\Bigr)\lambda_0=\bar\iota^{A'}\Bigl(v^e
 \bigl(D_e\ln\Omega\bigr)v_{A'A}\lambda^A\Bigr). \label{eq:3.3.9b}
\end{align}
Thus 
\begin{equation}
\Delta_{A'A}\lambda^A=v^e\bigl(D_e\ln\Omega\bigr)v_{A'A}\lambda^A,
\label{eq:3.3.10}
\end{equation}
i.e. $\Delta_{A'A}\lambda^A$ is proportional to the spinor field 
itself and the factor of proportionality is the normal directional 
derivative of the logarithm of the conformal factor. In the next 
section we will see that the boundary condition in the 
non-spherically symmetric case is a direct generalization of this. 

Conversely, any of (\ref{eq:3.3.9a}) and (\ref{eq:3.3.9b}) as a 
boundary condition implies that $\lambda_A$ is conformal to a spinor 
field which is constant in the flat 3-space. E.g. $0=\bar o^{A'}(v^e
(D_e\ln\Omega)v_{A'A}\lambda^A$ $-\Delta_{A'A}\lambda^A)=\Omega^2(
\hat{\edth}{}'\hat\lambda_0-\frac{1}{r}\hat\lambda_1)$, yielding the 
boundary condition for the constant spinor fields in flat 3-space.

\subsubsection{On solutions on the vacuum Brill--Lindquist/Bowen--York
data sets}
\label{sub-3.3.3}

The base manifold $\Sigma$ of the Brill--Lindquist data set is 
${\mathbb R}^3$ from which the points with the Cartesian coordinates 
$x^{\bi}_1$, \dots, $x^{\bi}_N$ have been removed. The conformal 
factor is $\Omega^{-1}(x^{\bi})=(1+\sum_{n=1}^N\frac{m_n}{2\vert x
^{\bi}-x^{\bi}_n\vert})^2$, where $m_1$, \dots, $m_N$ are positive 
constants; while the extrinsic curvature is vanishing. Under these 
conditions the vacuum constraint equations are satisfied \cite{BL}. 
(This data set is generalized by the Bowen--York data set, in which 
the extrinsic curvature is only traceless and is chosen to give a 
prescribed linear or angular momentum at spatial infinity \cite{BoY}. 
Since what is important for us is the conformal flatness of the 
3-metric and the vanishing of the trace of the extrinsic curvature, 
our analysis can be extended to this more general case without any 
difficulty.) This data set represents $N+1$ asymptotically flat ends 
at $r\rightarrow \infty$ and at the (missing) points $x^{\bi}_1$, 
\dots, $x^{\bi}_N$, and $N$ black holes with apparent horizons (in 
the form of minimal surfaces) surrounding the `internal' asymptotic 
ends $x^{\bi}_1$, \dots, $x^{\bi}_N$ (see also \cite{Gi72}). 

If $\lambda_A$ is a solution of the Witten equation on $(\Sigma,h
_{ab})$ such that it is bounded at the `outer' asymptotic end, i.e. 
when $r\rightarrow\infty$, then $\lambda_A=\Omega^{\frac{1}{2}}\hat
\lambda_A$, where $\hat\lambda_A$ is the sum of $N$ spinor fields 
$\hat\lambda^1_A$, \dots, $\hat\lambda^N_A$ whose components in the 
rescaled spin frame $\{\hat o^A,\hat\iota^A\}$ are given by 
(\ref{eq:3.2.9a})--(\ref{eq:3.2.9b}) with the centres at $x^{\bi}_1$, 
\dots, $x^{\bi}_N$, respectively. 
On the other hand, since $\Omega(x^{\bi})\rightarrow0$ as $\vert
x^{\bi}-x^{\bi}_n\vert^2$ when $x^{\bi}\rightarrow x^{\bi}_n$, the 
spinor field $\lambda_A$ is bounded near the `internal' asymptotic 
ends too precisely when it is asymptotically constant there. Such a 
spinor field is parametrized by $2(N+1)$ complex constants, but the 
solutions parametrized by constants belonging to a {\em proper 
submanifold} of $\mathbb{C}^{2(N+1)}$ have zeros.


\section{Boundary conditions for the constant and conformally 
constant spinor fields}
\label{sec-4}

\subsection{Boundary conditions for the constant spinors}
\label{sub-4.1}

Let $\Sigma$ be a subset of a spacelike hyperplane in Minkowski 
spacetime with a (not necessarily connected) smooth boundary ${\cal
S}$. Clearly, there are precisely two linearly independent spinor 
fields on $\Sigma$ which are constant in the sense that ${\cal D}_e
\lambda^A=0$. Then the spinor fields are constant on ${\cal S}$ with 
respect to $\Delta_e$, and hence, in particular, they satisfy $\Delta
_{A'A}\lambda^A=0$ too. Now we show that the converse is also true. 
Namely, {\em a solution $\lambda_A$ of the Witten equation is constant 
if and only if its restriction to the boundary satisfies the boundary 
condition $\Delta_{A'A}\lambda^A=0$}. 

The key ingredient is the (Reula--Tod form of the) Sen--Witten 
identity \cite{RT},
\begin{eqnarray}
-h^{ef}\bigl({\cal D}_e\lambda_A\bigr)\bigl({\cal D}_f\bar\lambda
 _{A'}\bigr)t^{AA'}\!\!\!\!&-\!\!\!\!&\frac{1}{2}t^a\,{}^4G_{ab}
 \lambda^B \bar\lambda^{B'}=2t^{AA'}\bigl({\cal D}_{AB'}\bar\lambda
 ^{B'}\bigr)\bigl({\cal D}_{A'B}\lambda^B\bigr)- \nonumber \\
\!\!\!\!&-\!\!\!\!&D_a\Bigl(t^{AB'}\bar\lambda^{A'}{\cal D}_{B'B}
 \lambda^B-t^{A'B}\bar\lambda^{B'}{\cal D}_{B'B}\lambda^A\Bigr),
\label{eq:4.1.1}
\end{eqnarray}
whose integral on $\Sigma$ can be written as 
\begin{eqnarray}
\Vert\lambda_A\Vert^2:\!\!\!\!&=\!\!\!\!&\int_\Sigma\Bigl(-h^{ef}
 \bigl({\cal D}_e\lambda_A\bigr)\bigl({\cal D}_f\bar\lambda_{A'}\bigr)
 t^{AA'}-\frac{1}{2}t^a\,{}^4G_{ab}\lambda^B\bar\lambda^{B'}\Bigr)
 {\rm d}\Sigma \label{eq:4.1.2} \\
\!\!\!\!&=\!\!\!\!&2\int_\Sigma t^{AA'}\bigl({\cal D}_{AB'}\bar
 \lambda^{B'}\bigr)\bigl({\cal D}_{A'B}\lambda^B\bigr){\rm d}\Sigma+
 \oint_{\cal S}\bar\lambda^{A'}\bar\gamma_{A'}{}^{B'}\Delta_{B'B}
 \lambda^B {\rm d}{\cal S}.  \nonumber
\end{eqnarray}
(Here ${}^4G_{ab}$ is the spacetime Einstein tensor, which, actually,
vanishes by assumption.) The first line of (\ref{eq:4.1.2}), which is 
essentially a Sobolev norm, is an integral of pointwise non-negative 
expressions (even if the data induced on $\Sigma$ were not flat, but 
satisfied the dominant energy condition), while the volume integral in 
the second line is vanishing by the Witten equation. Thus if $\lambda
^A$ is a solution of the Witten equation satisfying the boundary 
condition $\Delta_{A'A}\lambda^A=0$, then by (\ref{eq:4.1.2}) $\Vert
\lambda_A\Vert=0$, i.e. ${\cal D}_e\lambda^A=0$ follows.

If $\lambda_B$ is constant with respect to ${\cal D}_e$ on $\Sigma$, 
then its restriction to the boundary solves the 2-surface twistor 
equation as well: ${\cal T}_{A'AB}{}^C\lambda_C=0$. Although in the 
special cases considered in subsection \ref{sub-3.2} any of the two 
equations ${\cal T}_{A'AB}{}^C\lambda_C=0$ appears to be an 
appropriate boundary condition to single out the constant spinor 
fields, we do not have an equation analogous to (\ref{eq:4.1.2}) by 
means of which we could show a direct relationship between the 
${\cal D}_e\lambda_A$ derivative on $\Sigma$ and the 2-surface twistor 
derivative of $\lambda_A$ on ${\cal S}$. Thus, we formulate our 
boundary condition in terms of the 2-surface Dirac rather than the 
2-surface twistor operator (though the conformal invariance of the 
latter could have suggested to use it, especially in the conformally 
flat spaces). 

However, the boundary condition $\Delta_{A'A}\lambda^A=0$ appears to 
contradict the general theory of boundary value problems for elliptic 
systems: it is too strong, as it represents {\em two} rather than 
only one restriction on the spinor field. Although this boundary 
condition can be imposed in the special case of {\em flat} geometries, 
we should be able to reformulate it in a way that is compatible with 
the general theory of elliptic boundary value problems. 
In fact, the previous theorem can be proven with the following weaker, 
$\mathbb{R}$-linear (rather than $\mathbb{C}$-linear) boundary 
condition: for some complex function $\alpha:{\cal S}\rightarrow
\mathbb{C}$, which may depend on the spinor field, the spinor field 
satisfies $\Delta_{A'A}\lambda^A=\alpha\bar\gamma_{A'B'}\bar\lambda
^{B'}$. Or, in other words, it is required that the $\bar\gamma^{A'}
{}_{B'}\bar\lambda^{B'}$-component of the derivative of the spinor 
field vanishes: $\bar\lambda^{B'}\bar\gamma_{B'}{}^{A'}\Delta
_{A'A}\lambda^A=0$. This provides the correct number of boundary 
conditions, and can be rewritten as $\bar B^{A'}(\Delta_{A'A}\lambda
^A-\alpha\bar\gamma_{A'B'}\bar\lambda^{B'})=0$ for {\em any} spinor 
field $B^A$ and {\em some} $\alpha$ on ${\cal S}$.


\subsection{Conformally constant spinor fields on maximal, 
conformally flat hypersurfaces}
\label{sub-4.2}

By the results of subsections \ref{sub-2.1.1} and \ref{sub-4.1}, the 
solution $\lambda_A$ of the Witten equation (with conformal weight 
$w=-\frac{1}{2}$) is conformally constant in the conformally flat 
3-space precisely when $\hat\lambda^A:=\Omega^{-\frac{3}{2}}\lambda^A$ 
satisfies the boundary condition $\hat\Delta_{A'A}\hat\lambda^A=0$. 
Then by the conformal rescaling formula and $t^e\nabla_e\Omega=0$ this 
is equivalent to 
\begin{equation}
\Delta_{A'A}\lambda^A=\Bigl(\frac{1}{2}\delta_{A'A}\ln\Omega+v_{A'A}
v^eD_e\ln\Omega\Bigr)\lambda^A; \label{eq:4.2.1}
\end{equation}
i.e. the derivative $\Delta_{A'A}\lambda^A$ is proportional to the 
spinor field itself where the factor of proportionality is built 
from the derivatives of the conformal factor. (N.B.: On functions 
the derivative operator $\Delta_e$ coincides with the intrinsic 
Levi-Civita derivative operator $\delta_e$.) Contracting 
(\ref{eq:4.2.1}) with {\em any} given spinor field $B^A$ and 
denoting the coefficient of $\lambda^A$ in the resulting formula by 
$-A_A$, it has the form (\ref{eq:2.3.2}) with $f=0$. Moreover, 
assigning zero conformal weight to $B^A$, it is straightforward to 
check that, under a conformal rescaling $h_{ab}=\Omega^{-2}\hat h
_{ab}\mapsto\omega^2h_{ab}=\omega^2\Omega^{-2}\hat h_{ab}$ of the 
physical metric, the spinor field $A_A$ transforms just in the way 
required in the conformally invariant boundary condition 
(\ref{eq:2.3.6}). Indeed, since the concept of conformally constant 
spinor fields is conformally invariant, the boundary condition 
that specifies these should also be conformally invariant. 

Our aim is to characterize the conformally constant spinor fields 
among the solutions of the Witten equation by appropriate boundary 
conditions {\em in the physical, conformally flat} (rather than in 
the flat, rescaled) geometry of the hypersurface. By (\ref{eq:4.2.1}) 
the boundary condition must be searched for in the form 
\begin{equation}
\Delta_{A'A}\lambda^A=\Bigl(\frac{1}{2}\delta_{A'A}\alpha+v_{A'A}
\beta\Bigr)\lambda^A \label{eq:4.2.2}
\end{equation}
for some functions $\alpha,\beta:{\cal S}\rightarrow\mathbb{R}$. 
Clearly, for $\alpha={\rm const}$, $\beta=0$ this reduces to the 
boundary condition $\Delta_{A'A}\lambda^A=0$ for the constant 
spinors found in the previous subsection. However, these functions 
cannot be arbitrary, because the (globally conformally flat) 
geometry $(\Sigma,h_{ab})$ determines the conformal factor by 
means of which the geometry can be rescaled to be flat. On the 
other hand, there can be different flat 3-spaces that are globally 
conformal to each other (with non-trivial conformal factor). Hence 
$(\Sigma,h_{ab})$ does {\em not} determine the conformal factor 
completely. There can be some ambiguity in $\Omega$. Next we 
determine what kind of conditions should $\ln\Omega$ and $v^eD_e
\ln\Omega$ satisfy, and hence what kind of conditions should 
$\alpha$ and $\beta$ satisfy on ${\cal S}$. 

By the conformal rescaling formulae for the 3-dimensional Ricci 
tensor,
\begin{eqnarray}
0=\hat R_{ab}=R_{ab}\!\!\!\!&+\!\!\!\!&D_aD_b\ln\Omega-\bigl(D_a\ln
 \Omega\bigr)\bigl(D_b\ln\Omega\bigr)+ \nonumber \\
\!\!\!\!&+\!\!\!\!&h_{ab}\Bigl(D_eD^e\ln\Omega+\bigl(D_e\ln\Omega
 \bigr)\bigl(D^e\ln\Omega\bigr)\Bigr), \label{eq:4.2.3}
\end{eqnarray}
it is straightforward to calculate the 3-dimensional Einstein tensor. 
For its `constraint parts' on the boundary surface, $G_{ab}v^av^b$ 
and $G_{bc}v^b\Pi^b_a$, we obtain 
\begin{eqnarray}
&{}&\delta_e\delta^e\ln\Omega-\nu\bigl(v^eD_e\ln\Omega\bigr)-
 \bigl(v^eD_e\ln\Omega\bigr)^2=-G_{ab}v^av^b, \label{eq:4.2.4a} \\
&{}&\delta_a\bigl(v^eD_e\ln\Omega\bigr)-\bigl(v^eD_e\ln\Omega\bigr)
 \delta_a\ln\Omega-\nu_a{}^b\delta_b\ln\Omega=-G_{bc}v^b\Pi^c_a.
 \label{eq:4.2.4b}
\end{eqnarray}
Here $\nu_{ab}:=\Pi^c_a\Pi^d_bD_cv_d$, the extrinsic curvature of 
${\cal S}$ in $\Sigma$, and $\nu$ is its trace. Therefore, as we 
claimed, $\ln\Omega$ and $v^e(D_e\ln\Omega)$ in (\ref{eq:4.2.1}) 
are restricted by the geometry $(\Sigma,h_{ab})$ via 
(\ref{eq:4.2.4a})--(\ref{eq:4.2.4b}). On the other hand, we show 
that once $\ln\Omega$ or $v^e(D_e\ln\Omega)$ is given 
on ${\cal S}$, then the conformal factor on $\Sigma$ is already 
completely determined, provided Einstein's equations hold, ${}^4G
_{ab}=-\kappa T_{ab}$, and the weak energy condition $T_{ab}t^at^b
\geq0$ is satisfied. For, first observe that by the maximality of 
the hypersurface and the Hamiltonian constraint part of Einstein's 
equation $R=\chi_{ab}\chi^{ab}-\chi^2+2\kappa T_{ab}t^at^b\geq0$, 
i.e. the scalar curvature is non-negative. Second, by taking the 
trace of (\ref{eq:4.2.3}), it is straightforward to derive the {\em 
linear} equation 
\begin{equation}
D_eD^e\sqrt{\Omega}+\frac{1}{8}R\sqrt{\Omega}=0.
\label{eq:4.2.5}
\end{equation}
This equation is also known as the Lichnerowicz or Yamabe equation. 
Next suppose that $\Omega'$ is another solution of (\ref{eq:4.2.5}), 
and define $u:=\sqrt{\Omega}-\sqrt{\Omega'}$. Then $u$ also satisfies 
(\ref{eq:4.2.5}). Multiplying this by $u$ and integrating on $\Sigma$, 
by the Gauss theorem we obtain 
\begin{equation*}
\oint_{\cal S}\bigl(\sqrt{\Omega}-\sqrt{\Omega'}\bigr)v^aD_a\bigl(
\sqrt{\Omega}-\sqrt{\Omega'}\bigr){\rm d}{\cal S}=\int_\Sigma\Bigl(
-\bigl(D_au\bigr)\bigl(D^au\bigr)+\frac{1}{8}Ru^2\Bigr){\rm d}\Sigma,
\end{equation*}
where both terms in the integrand on the right-hand side are 
pointwise non-negative. Thus, if either $\Omega$ and $\Omega'$, or 
$v^eD_e\sqrt{\Omega}$ and $v^eD_e\sqrt{\Omega'}$ coincide on ${\cal S}$, 
then $\Omega'=\Omega$ on the whole $\Sigma$, too. Therefore, $\Omega$ 
is, in fact, completely determined e.g. by {\em its own value on the 
boundary}, and the ambiguity in $\Omega$ corresponds to the 
non-uniqueness of the solution of (\ref{eq:4.2.4a})--(\ref{eq:4.2.4b}). 

To determine this ambiguity, let us suppose that both $\Omega$ and 
$\Omega':=\omega^{-1}\Omega$ are solutions of (\ref{eq:4.2.3}). Then 
the difference of this equation for $\Omega$ and for $\Omega'$ is a 
differential equation, whose trace (multiplied by $\Omega$) and 
trace-free part (multiplied by $2\Omega^{-2}$) yield the 
system of equations 
\begin{eqnarray}
&{}&D_e\Bigl(\Omega D^e\omega^{-\frac{1}{2}}\Bigl)=0,
 \label{eq:4.2.6a}\\
&{}&D_a\bigl(\Omega^{-2}D_b\omega\bigr)+D_b\bigl(\Omega^{-2}D_a
 \omega\bigr)-\frac{2}{3}h_{ab}D_e\bigl(\Omega^{-2}D^e\omega\bigr)=0.
 \label{eq:4.2.6b}
\end{eqnarray}
However, the second is just the conformal Killing equation for the 
{\em hypersurface--orthogonal} vector field $K_a:=\Omega^{-2}D_a
\omega$. Thus by determining those conformal Killing vectors $K_a$ 
for which $\Omega^2K_a$ is a gradient, we obtain a class of 
functions $\omega$, and the ambiguity in the conformal factor is 
represented by those of these functions that solve (\ref{eq:4.2.6a}), 
too. It might be worth noting that in the Cartesian coordinates $\{
x^{\bi}\}$ of the {\em flat} 3-space $(\Sigma,\hat h_{ab})$ the 
solution $\omega$ can be given explicitly. In fact, in terms of the 
rescaled (flat) metric (\ref{eq:4.2.6a}) and (\ref{eq:4.2.6b}) take 
the form 
\begin{equation*}
\hat D_e\hat D^e\omega^{-\frac{1}{2}}=0, \hskip 20pt
\hat D_a\hat D_b\omega-\frac{1}{3}\hat h_{ab}\hat D_e\hat D^e\omega
=0.
\end{equation*}
Then the solution of these equations is $\omega=C\delta_{\bi\bj}
(x^{\bi}+C^{\bi})(x^{\bj}+C^{\bj})$ for some constants $C$ and $C
^{\bi}$. 

Thus, to summarize, the boundary condition that specifies the 
conformally constant spinor fields among the solutions of the Witten 
equation is (\ref{eq:4.2.2}), where the functions $\alpha$ and 
$\beta$ are solutions of the differential equations 
\begin{eqnarray}
&{}&\delta_e\delta^e\alpha-\nu\beta-\beta^2=-G_{ab}v^av^b,
 \label{eq:4.2.7a} \\
&{}&\delta_a\beta-\beta\delta_a\alpha-\nu_a{}^b\delta_b\alpha=-G_{bc}
 v^b\Pi^c_a. \label{eq:4.2.7b}
\end{eqnarray}
Since $\ln\Omega$ and $v^e(D_e\ln\Omega)$ solve these, we know that 
such $\alpha$ and $\beta$ exist; i.e. our boundary condition can 
always be imposed. The solution determines a conformal factor on 
$\Sigma$ in a unique way such that the corresponding conformal 
rescaling yields a flat 3-space and a constant spinor field. Although 
the solution $(\alpha,\beta)$ of (\ref{eq:4.2.7a})--(\ref{eq:4.2.7b}) 
is not unique, the ambiguity is completely controlled by the function 
$\omega$, which solves (\ref{eq:4.2.6a})--(\ref{eq:4.2.6b}). The 
meaning of this $\omega$ is, however, only a `pure gauge', telling us 
which flat 3-metric is chosen to be the `reference' with respect to 
which the physical metric is conformally flat.


\section{Summary and conclusions}
\label{sec-5}

In certain physical problems and in the study of the structure of 
the field equations it is useful to reduce the gauge freedom 
of the theory by some appropriate gauge condition. On a spacelike
hypersurface this could be the use of the orthonormal triad field 
coming from the spinor fields solving the Witten equation, or the 
triad field satisfying the so-called special orthonormal frame 
gauge condition. These conditions take the form of some elliptic 
partial differential equations, thus their solutions can be 
controlled by the boundary condition. However, for general boundary 
conditions the triad fields (either built from the solution of 
the Witten equation or satisfying the frame gauge condition) can 
be degenerate. Thus the proper gauge condition should consist of 
the elliptic p.d.e. {\em and} the boundary conditions selecting the 
{\em globally nonsingular ones} from the infinitely many solutions. 

In the present paper these boundary conditions were investigated on 
maximal, globally intrinsically conformally flat spacelike 
hypersurfaces. (Such hypersurfaces are e.g. the Reissner--Nordstr\"om, 
the Brill--Lindquist and the Bowen--York initial data sets for 
finitely many black holes in asymptotically flat spacetimes.) 
On such hypersurfaces the two gauge conditions above are equivalent, 
and hence can be studied simultaneously. We determined the boundary 
conditions that characterize (1) the constant spinor fields on 
compact domains in intrinsically and extrinsically flat hypersurfaces, 
and (2) the conformally constant spinor fields on compact domains 
in maximal, intrinsically globally conformally flat spacelike 
hypersurfaces (provided the Hamiltonian constraint holds and the weak 
energy condition is satisfied). Thus, in particular, the special 
orthonormal frame gauge condition can always be satisfied by globally 
non-degenerate frames on arbitrary compact domains in such 
hypersurfaces. 

The exact solutions of subsections of \ref{sub-3.2}--\ref{sub-3.3} 
show that many of the `natural' boundary conditions (appearing in 
various special problems) yield a degenerate triad field. In 
addition, the example of the Reissner--Nordstr\"om (and the more 
general Brill--Lindquist and Bowen--York) data sets show that there 
is a (still not quite well understood) interplay between the 
boundary conditions, the global topology of the hypersurface and 
the existence/non-existence of zeros of the solutions of the Witten 
equation: the boundary conditions on the different connected 
components of the boundary cannot be chosen independently.


\section{Acknowledgments}

J.F. is grateful to the Research Institute for Particle and Nuclear 
Physics, Budapest, and to the Albert Einstein Institute, Golm, for 
kind hospitality while this work was finished. The work of J.M.N. 
was supported by the National Science Council of the R.O.C. under 
the grant number NSC-99-2112-M-008-004 and in part by the Taiwan 
National Center of Theoretical Sciences (NCTS). L.B.Sz. is grateful 
to the Center for Mathematics and Theoretical Physics, National 
Central University, Chung-Li, Taiwan, for hospitality during the 
preparation of a preliminary version of this work. This work was 
partially supported by the Marsden Fund of the Royal Society of New 
Zealand under grant number UOO0922 and the Hungarian Scientific 
Research Fund (OTKA) grant K67790.


\begin{thebibliography}{999}

\bibitem{GH}
          G. W. Gibbons, S. W. Hawking, G. T. Horowitz, M. J. Perry,
           Positive mass theorem for black holes, Commun. Math. Phys.
           {\bf 88} 295--308 (1983)
\bibitem{RT}
          O. Reula, K. P. Tod, Positivity of the Bondi energy, J. Math.
           Phys. {\bf 25} 1004--1008 (1984)

\bibitem{Ne}
         J. M. Nester, A positive gravitational energy proof, Phys.
         Lett. A {\bf 139} 112-114 (1989)
\bibitem{N1}
         J. M. Nester, A gauge condition for orthonormal three-frames,
          J. Math. Phys. {\bf 30} 624--626 (1989)
\bibitem{N2}
         J. M. Nester, Special orthonormal frames, J. Math. Phys.
          {\bf 33} 910--913 (1992)

\bibitem{J}
          J. Frauendiener, Triads and the Witten equation, Class.
           Quantum Gravity, {\bf 8} 1881--1187 (1991)
\bibitem{DMH}
         A. Dimakis, F. M\"uller-Hoissen, On a gauge condition for
          orthonormal three-frames, Phys. Lett. A {\bf 142} 73--74
          (1989)

\bibitem{bar} C. B\"ar, On nodal sets for Dirac and Laplace operators,
          Commun. Math. Phys. {\bf 188} 709--721 (1997)
          
\bibitem{Ne07} J. M. Nester, On the zeros of spinor fields and an 
          orthonormal frame gauge condition, in {\em Proceedings of 
          the Eleventh Marcel Grossmann Meeting on General Relativity} 
          (Berlin, Germany, 23--29 July 2006) pp. 1332--1334, eds. H. 
          Kleinert, R. T. Jantzen, World Scientific, Singapore 2007

\bibitem{Gi} G. W. Gibbons, The isoperimetric and Bogomolny
          inequalities for black holes, in {\it Global Riemannian
          Geometry}, Ellis Horwood Series in Mathematics and Its
          Applications, Ed. T. J. Willmore and N. J. Hitchin,
          pp. 194--202, Ellis Horwood Ltd, Chichester 1984

\bibitem{BL} D. R. Brill, R. W. Lindquist, Interaction energy in
          geometrostatics, Phys. Rev. {\bf 131} 471-476 (1963)

\bibitem{BoY} J. M. Bowen, J. W. York, Time-asymmetric initial data for
          black holes and black hole collisions, Phys. Rev. D {\bf 21}
          2047-2056 (1980)

\bibitem{PR}
          R. Penrose, W. Rindler, {\it Spinors and Spacetime}, Vol 1,
          and Vol 2, Cambridge University Press, Cambridge 1984 and
          1986

\bibitem{HT}
          S. A. Hugget, K. P. Tod, {\it An Introduction to Twistor
           Theory}, vol 4 of London Mathematical Society Student
           Texts, Cambridge University Press, Cambridge 1985

\bibitem{Se}
         A. Sen, On the existence of neutrino `zero-modes' in vacuum
          spacetimes, J. Math. Phys. {\bf 22} 1781--1786 (1981)

\bibitem{Sz1}
          L. B. Szabados, Two-dimensional Sen connections in
           general relativity, Class. Quantum Grav. {\bf 11}
           1833--1846 (1994)
\bibitem{D} S. Dain, Elliptic systems, in {\it Analytical and
          Numerical Approaches to Mathematical Relativity}, Lecture
          Notes in Physics vol 692, pp. 117-139, Ed. J. Frauendiener,
          D. J. W. Giulini and V. Perlick, Springer Berlin/Heidelberg
          2006, gr-qc/0411081

\bibitem{DM}
          A. J. Dougan, L. J. Mason, Quasilocal mass constructions
          with positive energy, Phys. Rev. Lett. {\bf 67} 2119--2122
          (1991)
\bibitem{Sz2}
          L. B. Szabados, Two-dimensional Sen connections and
           quasi-local energy-momentum, Class. Quantum Grav. {\bf 11}
           1847--1866 (1994)
\bibitem{Sz3}
          L. B. Szabados, On certain quasi-local spin-angular momentum
          expressions for large spheres near the null infinity, Class.
          Quantum Grav. {\bf 18} 5487--5510 (2002),
          Corrigendum:  Class. Quantum Grav. {\bf 19} 2333 (2003)

\bibitem{Gi72} G. W. Gibbons, The time symmetric initial value problem
          for black holes, Commun. Math. Phys. {\bf 27} 87--102 (1972)



\end{thebibliography}
\end{document}